\documentclass[sigconf]{acmart}

\usepackage{multirow}
\usepackage{makecell,acmart-taps}
\usepackage{booktabs}
\usepackage{array}
\newcolumntype{P}[1]{>{\raggedright\arraybackslash}p{#1}}
\usepackage{makecell}

\newlength{\cuaindent}
\usepackage{pifont,xcolor,framed}

\definecolor{shadecolor}{rgb}{0.82,0.82,0.82}

\AtBeginDocument{%
  }

\copyrightyear{2026}
\acmYear{2026}
\setcopyright{cc}
\setcctype{by}
\acmConference[CHI '26]{Proceedings of the 2026 CHI Conference on Human Factors in Computing Systems}{April 13--17, 2026}{Barcelona, Spain}
\acmBooktitle{Proceedings of the 2026 CHI Conference on Human Factors in Computing Systems (CHI '26), April 13--17, 2026, Barcelona, Spain}
\acmDOI{10.1145/3772318.3791376}
\acmISBN{979-8-4007-2278-3/2026/04}

\newcommand{\revised}[1]{#1}

\begin{document}

\title{Co-Designing Multimodal Systems for Accessible Asynchronous Dance Instruction}

\author{Ujjaini Das}
\affiliation{
  \institution{University of Texas, Austin}
  \country{Austin, Texas, USA}}
\email{ujjaini@utexas.edu}

\author{Shreya Kappala}
\affiliation{
  \institution{University of Texas, Austin}
  \country{Austin, Texas, USA}}
\email{shkap13@utexas.edu}

\author{Meng Chen}
\affiliation{
  \institution{University of California, Berkeley}
  \country{Berkeley, California, USA}}
\email{meng.chen@berkeley.edu}

\author{Mina Huh}
\affiliation{
  \institution{University of California, Berkeley}
  \country{Berkeley, California, USA}}
\email{minahuh@berkeley.edu}

\author{Amy Pavel}
\affiliation{
  \institution{University of California, Berkeley}
  \country{Berkeley, California, USA}}
\email{amypavel@eecs.berkeley.edu}


\begin{abstract}
Videos make exercise instruction widely available, but they rely on visual demonstrations that blind and low vision (BLV) learners cannot see. While audio descriptions (AD) can make videos accessible, describing movements remains challenging as the AD must convey \textit{what} to do (mechanics, location, orientation) and \textit{how} to do it (speed, fluidity, timing). Prior work thus used \textit{multimodal instruction} to support BLV learners with individual simple movements. However, it is unclear how these approaches scale to dance instruction with unique, complex movements and precise timing constraints. To inform accessible \revised{asynchronous} dance instruction systems, we conducted three co-design workshops (N=28) with BLV dancers, instructors, and experts in sound, haptics, and AD. Participants designed 8 systems revealing common themes: staged learning to dissect routines, crafting vocabularies for movements, and selectively using modalities—narration for movement structure, sound for expression, and haptics for spatial cues. We conclude with design \revised{implications} to make learning dance accessible. 
\end{abstract}






\maketitle

\section{Introduction}

\revised{Millions of people now participate in exercises such as dance, yoga, and fitness using online videos. 
Exercise videos allow people to learn on their own time and in their own environment. Thus, their popularity expanded with the COVID-19 pandemic~\cite{sui2022engagement, delabary2022online, suzuki2022dance, chen2020coronavirus} and the corresponding rise in the use of video platforms~\cite{Mitkina2023youtube, nuhn2025tiktok, warburton2024tiktok, hong2020youtube}.
Such videos also have the potential to be particularly useful for people with disabilities as remote learning reduces travel barriers associated with in-person physical instruction~\cite{jaarsma2014barriers, lee2014calibration}, and increased physical activity can improve life satisfaction~\cite{labudzki2013lifesatisfaction}. Yet, this revolution in accessible fitness has systematically excluded blind and low vision (BLV) learners, as exercise videos rely on visual demonstrations~\cite{desilva2023understanding} that BLV learners cannot see. Compared to synchronous settings, asynchronous learning from online videos poses challenges to BLV learners given the lack of feedback and instructor adaptation~\cite{jiang2025audio}.}

Audio descriptions (AD) can make videos accessible by narrating the visual content that is important to understand the video. While AD is now common for TV shows and movies~\cite{jordan2018media}, it remains virtually non-existant for user-generated videos on platforms such as YouTube~\cite{liu2021makes} and TikTok~\cite{van2024making}. Thus, prior work explored manual~\cite{YouDescribe, wang2025cosight}, semi-automated~\cite{pavel2020rescribe,liu2022crossa11y}, and automated~\cite{van2024making,wang2021automaticAD} approaches to adding audio descriptions to videos. But, complex movements that appear in exercise videos remain challenging to describe with verbal narration alone~\cite{desilva2023understanding} as the AD must convey both \textit{what} to do (complex body positions, spatial location, and orientation) and \textit{how} to do it (speed, fluidity, pacing, and expressiveness). Even if the \textit{instructions} of the video are understandable, it remains challenging for BLV audience members who can not visually compare their own progress to gain feedback~\cite{huh2025vid2coach}.

Prior work thus explored how to provide accessible \textit{feedback} to BLV learners with multi-modal approaches --- from verbal feedback to correct yoga pose positioning~\cite{rector2013yoga} to haptic and audio feedback to improve performance in tennis~\cite{morelli2010vitennis} and bowling~\cite{morelli2010vibowling} exergames. 
Such tools provide \textit{feedback loops} that can improve performance and confidence~\cite{nikooyan2015reward, todorov1997augmented}, and alternative modalities (sound, haptic, tactile) that can support a more complete understanding of movements when combined with verbal narrations~\cite{blasing2021dance, seham2015extending, artpradid2023kinestheticEmpathicWitnessing}. 
Dance movements in particular can be challenging to convey and correct due to the diverse, complex positioning of the body, such that prior work explored how to make movement instruction more accessible for in-person Contemporary dance classes with multiple modalities~\cite{desilva2025sensing}. While this rich body of prior work demonstrates that communicating and correcting individual movements to BLV learners is possible, it is not clear how to scale such knowledge to long, multi-move routines of online videos without the support of co-located human instruction and feedback. \revised{Our work explores creating systems to translate \textit{static instructions} from asynchronous videos into accessible \textit{instruction systems} for BLV learners.}

We use a \textit{participatory design approach} to investigate how to make \revised{asynchronous} dance instruction accessible to BLV learners by including BLV learners and dance instructors as equal designers rather than mere consumers of technology. \revised{As we aim to maintain the advantages of asynchronous learning from dance tutorial videos available online, we define the solution that our work explores as \textbf{asynchronous dance instruction systems} for BLV learners.} 
We select dance as our domain of interest due to the rich complexity in movement mechanics, timing, and multi-movement sequences that makes dance particularly challenging to learn. 
To co-design asynchronous dance instruction systems, we conduct \textit{co-design workshops} with participants that have a wide range of relevant expertise in learning, instructing, and communicating around dance: BLV dance learners, dance instructors, technical experts (in AD, sound, and haptics), and participants with multiple types of expertise (\textit{e.g.}, BLV dance instructors). Our workshops aimed to answer the research question: \textit{How should we design systems to support asynchronous dance instruction for BLV learners?}

Over the course of three in-person co-design workshops with eight design groups, 28 total participants generated considerations for accessible asynchronous dance instruction systems and created eight unique system designs. Our analysis revealed several common themes across these system designs: using staged learning to disentangle placement and timing concerns, breaking routines into small movements, crafting multimodal vocabularies for movements (\textit{e.g.}, unique names, or sound cues), and selectively using multiple modalities across learning stages --- \textit{e.g.}, narration for movement mechanics, sound for conveying expression, and haptics for spatial cues.
\revised{We conclude with system architectures and design implications surfaced through co-design for accessible asynchronous dance learning systems to inform future system design and research.}


\section{Related Work}
Our work relates to prior work on video accessibility, accessible movement instruction, and multimodal feedback for accessibility.

\subsection{Video Accessibility}
Videos are a powerful medium for learning new skills, enabling learners to engage on their own schedule, replay segments, and control the pace~\cite{chang2021rubyslippers, huh2025vid2coach}.
There is a plethora of dance educational videos online, yet BLV learners find it challenging to utilize them with underdescribed visuals. For instance, narration and on-screen text rarely describe what bodies are doing in sufficient detail for learning~\cite{liu2021makes, liu2022crossa11y, peng2021say}. 
Existing practice relies on \textit{audio description} authored by volunteers~\cite{YouDescribe} or professionals~\cite{pavel2020rescribe}, but authoring is time-consuming and coverage is limited. While recent systems explore automatic descriptions by identifying scene changes~\cite{huh2023avscript, chang2024worldscribe} or providing interactive access~\cite{van2024making, ning2024spica, huh2022cocomix}, these tools primarily target high-level understanding in entertainment or general video contexts rather than conveying fine-grained, continuous motion needed for learning dance.

Beyond high-level AD, recent work has explored how-to video systems for BLV learners that scaffold procedural tasks (\textit{e.g.,} cooking, crafts) by segmenting actions into discrete steps, aligning guidance to natural pauses, and providing correctness checks or alerts~\cite{huh2025vid2coach, li2025oscar, ning2025aroma}. 
However, dance features fluid transitions, overlapping actions across multiple body parts, and values expressing intent beyond right–wrong verification~\cite{desilva2023understanding, ravn2017dancing}. 
In this work, we focus on the underexplored setting of dance learning videos and investigate multimodal approaches that complement AD to make dance instruction more intuitive and efficient for BLV learners.

\subsection{Accessible Movement Instruction} 
Prior work explored movement instruction to BLV learners in in-person classes and game-like training tools.
In-person movement learning has emphasized embodied, multi-sensory pedagogy: pairing verbal guidance with physical in one-to-one settings to convey orientation, timing, and movement qualities that verbal description alone underspecifies~\cite{seham2015extending,desilva2023understanding, aggravi2016hapticskiing}. Workshops in contemporary dance further explore sound cues and haptics to support spatialization and improvisation in dance~\cite{desilva2025sensing}. These approaches are effective in the studio but typically presume co-located instruction with touch, limiting remote access. 

Exergames, or video games that integrate physical exercises into engaging experiences, often combine modalities that can increase engagement without vision. For instance, audio+haptic tennis~\cite{morelli2010vitennis} and bowling~\cite{morelli2010vibowling} have been shown to improve guidance and enjoyment.
Prior works have also explored sensor-based systems for \textit{home exercise}~\cite{malik2021aerobic, rector2013yoga}, yet they focus on guiding stance and alignment by evaluating \textit{discrete} poses, rather than the continuous transitions or expression central to dance. Feedback is typically stepwise (\textit{e.g.,} verbal prompts, success tones) and correctness is judged at end-states (\textit{e.g.,} hit/hold/balance), not on how motions unfold or whether movement \emph{qualities} align with intent. 
Dance learning from video, by contrast, demands precise timing, nuanced transitions between shapes, and attention to expressive dynamics. 
Our work addresses this gap by exploring accessible dance education in ~\textit{asynchronous} video settings, which present unique opportunities and considerations for multimodal solutions.

\subsection{Multi-Modal Feedback for Accessibility}
Across and beyond accessibility domains, combining modalities has often been shown to provide advantages over single-channel feedback. 
In accessibility research, multimodal feedback has been used to guide both visual media consumption (\textit{e.g.,} videos~\cite{ning2024spica}, presentations~\cite{peng2021slidecho}) and creation tasks (\textit{e.g.,} artboards~\cite{10.1145/3526114.3558695}, video editing~\cite{huh2023avscript}, DIY assembly~\cite{10.1145/3586182.3616646}) for BLV people.
In museum contexts, pairing verbal description with non-verbal audio effectively communicates both objective detail and affect~\cite{rector2017eyesfreeart}. For dance, this can support rich information delivery: conveying objective elements (\textit{e.g.,} timing, sequencing, spatial targets) as well as expressing higher-level movement qualities (\textit{e.g.,} fluidity, energy, lightness) that non-verbal audio can carry~\cite{frid2018motionfluidity, landry2020interactive}. As Blasing et al.~\cite{blasing2021dance} highlights ~\textit{``the more modalities through which an observer has experienced a movement, the more enjoyment the observer derives from watching that movement''.} However, most prior work emphasizes observational experiences. Our work focuses on education, exploring how multimodal cues can complement and strengthen each other to make dance learning more precise and expressive for BLV learners.
\begin{figure}
    \centering
    \includegraphics[width=3.33in]{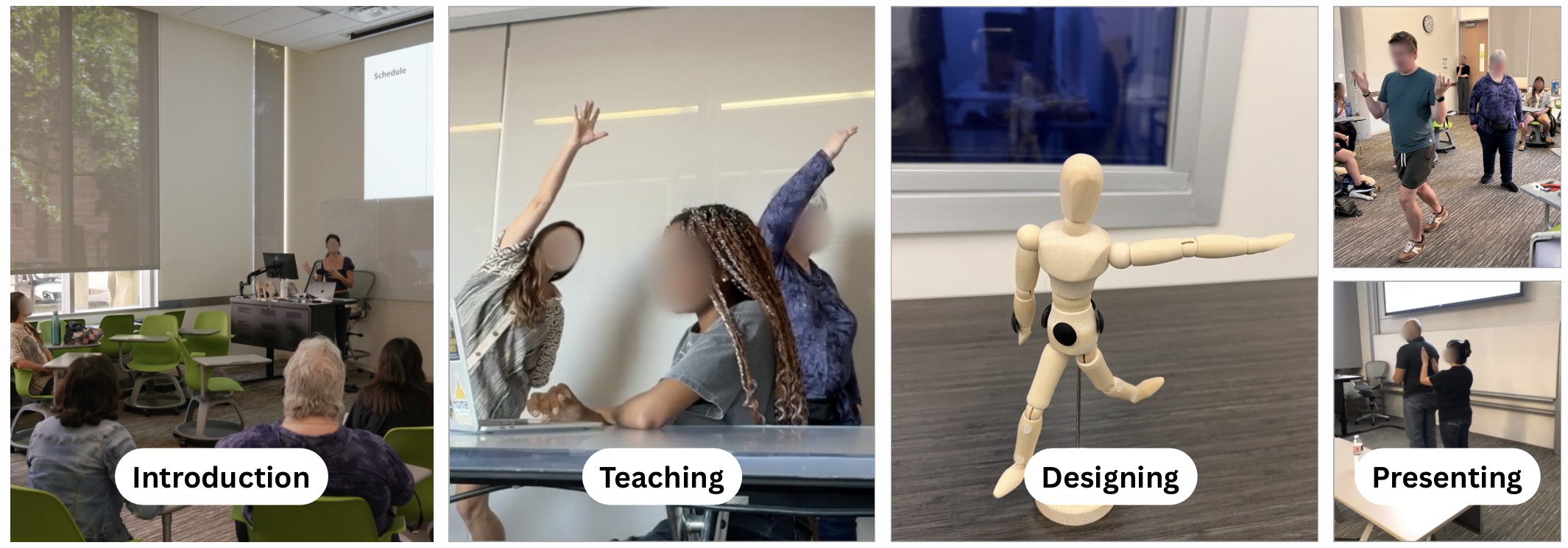}
    \caption{Our co-design workshops featured 4 major phases. The participants were informed of the workshop goals and their task (Introduction), then they split into smaller groups to teach the dance clips to one another (Teaching), design a system to effectively teach the clip using what they learned from teaching one another (Designing) and then shared their systems with all the other groups (Presentation). }
    \label{fig:workflow}
\end{figure}
\section{Method}

We aimed to answer the following research questions:

\begin{itemize}
    \item[\textbf{RQ1}] \revised{What are the current needs and design considerations of BLV learners seeking to learn dance asynchronously?}
    \item[\textbf{RQ2}] \revised{How can we design systems to make asynchronous dance instruction accessible through verbal narration, non-verbal audio, and/or haptics?}
\end{itemize}
\noindent To incorporate the deep expertise and lived experience of people with disabilities and relevant experts, we selected to use \textbf{co-design workshops} --- a participatory design approach that invites community members (\textit{e.g.}, BLV learners, dance instructors) to be creators and originators of the technology that they use rather than only end consumers~\cite{kane2014collaboratively, Magnusson2018codesigningVI}. 
Similar to prior co-design workshops~\cite{Magnusson2018codesigningVI, aflatoony2020ATmakers, valencia2021codesign}, we aimed to incorporate perspectives of not only potential end consumers of the technology (BLV dance learners) but also relevant experts in dance instruction (BLV and sighted dance instructors), and experts in multimodal technologies (audio description, sound design, haptics) to build on their complementary perspectives.

\subsection{Workshop Design}

\revised{We held three separate in-person co-design workshops that all featured the same five stages} (Figure~\ref{fig:workshopdesign}): (1) a pre-workshop \textbf{survey} to surface relevant individual dance backgrounds and lived experiences, (2) a large group \textbf{introduction} to share survey findings and discuss modalities for dance instruction, (3) small groups \textbf{teaching} example dances using non-visual techniques, then (4) \textbf{designing} systems for non-visual dance instruction using provided prototyping materials, and (5) \textbf{presenting} their systems to back to the large group for discussion. The four in-person workshop phases are shown in Figure~\ref{fig:workflow}. 
Each workshop was three hours long, and we compensated each participant \$90 for their participation. This study was approved by our institution's Institutional Review Board (IRB). 

\subsubsection{{\bf Participants.}} We began recruitment by reaching out to dance professors and students with teaching experience at our institution. We additionally connected with local BLV communities, and freelance sound producers and dance instructors. Furthermore, we partnered with an arts and accessibility organization and brought in two of their directors. Our three workshops included a total of 28 participants (16 female, 11 male, 1 non-binary), including 13 BLV participants (B1-B8b), 11 dance teachers (D1-D8b), two sound designers (S2, S6), a haptics engineer (H1), and an audio describer (A8) (Table~\ref{tab:participants}). \revised{Many participants had combined expertise: 10 of 13 BLV participants had prior dance experience, B3b, D4a and H1 also had sound design expertise, and B2a, B3a, B7b, and B8a also had dance instruction expertise.}

\begin{table*}[h!]
\centering
\resizebox{\textwidth}{!}{%
\renewcommand{\arraystretch}{1.3} %
\begin{tabular}{l l l l l l l l l}
\toprule
\textbf{ID} & \textbf{Gender} & \textbf{Age} & \textbf{Vision} & \textbf{Onset} & \textbf{Occupation} & \textbf{Skill Set} & \textbf{Years} & \textbf{Dance Styles} \\
\midrule
B1 & Female & 43 & Totally blind & 2 & Technical support & D & $<1$ & Ballroom dancing \\
B2a & Female & 66 & Legally blind & 6 & Production worker & D/DI & 10 & Ballet, jazz, modern \\
B2b & Male & 42 & Totally blind & 21 & Production worker & – & – & – \\
B3a & Female & 47 & Legally blind & 15 & Staff trainer & D/DI & 40 & Hip hop, country, salsa, ballet \\
B3b & Male & 53 & Totally blind & 23 & Musician & S & 41 & - \\
B4 & Female & 37 & Totally blind & 0 & Student & D & $<1$ & Partner dancing \\
B5a & Female & 40 & Legally blind & 3 & Teacher & D & 2 & Ballroom dancing \\
B5b & Female & 30 & Totally blind & 0 & Data entry & – & – & – \\
B6 & Female & 45 & Legally blind & 20 & Sales clerk & D & 40 & Latin dances \\
B7a & Male & 47 & Totally blind & 18 & Warehouse specialist & D & 4 & Hip-hop, country, line dancing \\
B7b & Male & 60 & Legally blind & 20 & Unemployed & D/DI & 40 & Ballet, ballroom, tap \\ 
B8a & Female & 38 & Legally blind & 6 & Admissions & D/DI & 8 & Partner dancing \\
B8b & Male & 44 & Legally blind & 20 & Public servant & D & 0.5 & Partner dancing \\
\midrule
D1 & Female & 28 & Sighted &  & Tax volunteer & D/DI & 5 & Bharatnatyam \\
D2 & Female & 21 & Sighted &  & Student & D/DI & 16 & Ballet, modern, hip-hop, African \\
D3 & Female & 20 & Sighted &  & Student & D/DI & 2 & Hip-hop, street styles \\
D4a & Female & 43 & Sighted &  & Dance teacher & D/DI, S & 23 & Ballet, contemporary, improvisation \\
D4b & Female & 20 & Sighted &  & Desk assistant & D/DI & 2 & Hip-hop, modern \\
D5 & Male & 38 & Sighted &  & Dance teacher & D/DI & 10 & Ballet \\
D6 & Male & 38 & Sighted &  & Professor & D/DI & 15 & Ballroom dancing \\
D7a & Non-binary & 20 & Sighted &  & Student & D/DI & 7 & Hip-hop, kpop \\
D7b & Female & 45 & Sighted &  & Director of Dance & D/DI & 20 & Contemporary \\
D8a & Female & 21 & Sighted &  & Student & D/DI & 5 & Contemporary \\
D8b & Male & 19 & Sighted &  & Student & D/DI & 3 & Contemporary \\ \midrule
H1 & Male & 22 & Sighted &  & Self-employed & S, H & 6 S & - \\
S2 & Male & 29 & Sighted &  & Audio producer & S & 7 & - \\
S6 & Male & 42 & Sighted &  & Audio producer & S & 20 & - \\
A8 & Female & 74 & Sighted &  &  Audio Describer & A & 20 & - \\
\bottomrule
\end{tabular}
}
\caption{Participant demographic and background information, including gender, age, current vision, age of onset if blind, occupation, skill set (D = Dance, DI = Dance instruction, S = Sound design, H = Haptics engineering, A = Audio description), years of experience of the skill set, and dance styles if prior dance experience. \revised{Participant IDs are prefixed with ``B'' for BLV participants, or by primary expertise area if sighted. Numbers in participant IDs indicate the group number that the participant was a part of during \textit{Teaching} and \textit{Designing} stages.}}
\label{tab:participants}
\end{table*}

\subsubsection{{\bf Workshop Stages}} 
\begin{figure*}[t]
    \centering
    \includegraphics[width=\linewidth]{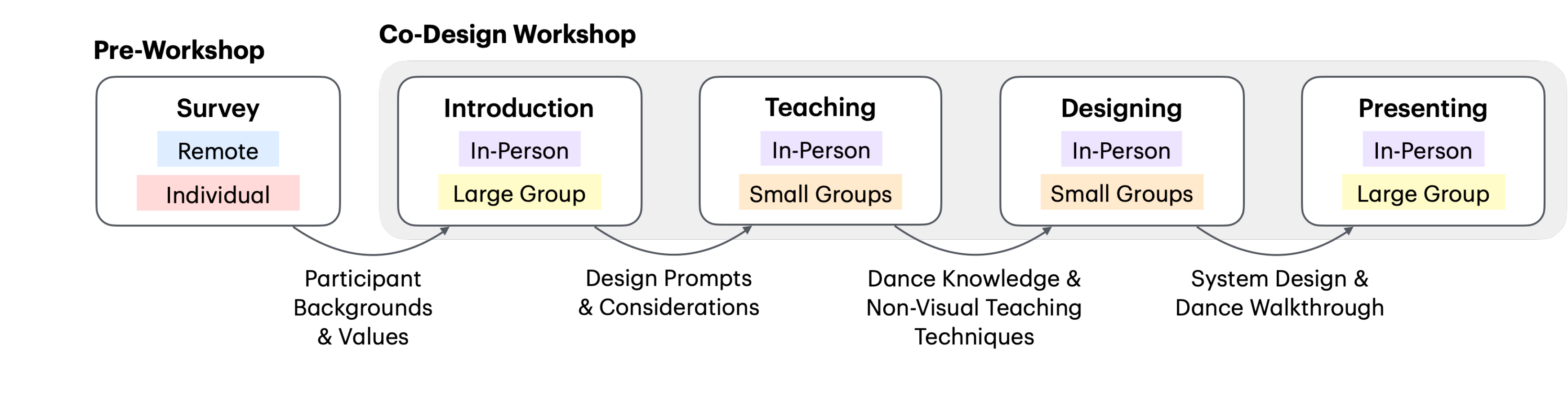}
    \caption{\revised{Each individual workshop followed a five-stage process: 1) pre-workshop survey, 2) co-design workshop introduction, 3) teaching, 4) designing, and 5) presenting.}}
    \label{fig:workshopdesign}
\end{figure*}

We designed the workshop stages (Table \ref{fig:workshopdesign}) to maximize contributions by all workshop members, including using accessible workshop best practices (\textit{e.g.}, all materials accessible, ask for accommodations in advance, provide multiple opportunities to update access requests) and using a mix of individual, small group, and large group activities. We structured large group activities to encourage participation from every workshop member (\textit{e.g.}, think-pair-share, explicit turn order) and we selected small groups \revised{of 3-5 participants each (Table~\ref{tab:groups})} to represent diverse expertise to foster complementary contributions. \\ 

\noindent \textit{\textbf{Pre-Session Survey.}} We invited BLV participants (13 participants) and dance instructors (11 participants) to share prior experiences with dance instruction in a pre-session survey. We asked BLV participants about their prior experiences learning movement skills, approaches for accessing remote movement education, and any benefits and challenges of remote instruction. We asked dance instructors about their teaching techniques, changes for remote settings and BLV settings (if applicable), and any observed or anticipated challenges of non-visual instruction. BLV dance instructors were asked both sets of questions. 


From the survey results and our review of prior literature~\cite{blasing2021dance, desilva2023understanding, desilva2025sensing} we distilled three main themes for accessible \revised{asynchronous} dance learning to address during the co-design workshop: (1) \textbf{clarity}, as visual demonstrations are often not fully supplemented by other sensory forms, (2) \textbf{feedback}, as remote learners are currently unable to have their movements corrected by visual comparison or touch, and (3) \textbf{temporal and emotional qualities}, as existing verbal descriptions insufficiently convey fluidity, weight, timing, emotions, and textures in movements that are important for learning dance. \\ 

\noindent \textit{\textbf{Introduction: 40 minutes.}} We started each workshop with an interactive presentation with group discussions to build participant familiarity and prime considerations for the following small-group design exercises. 

We first asked all participants and research staff to take turns introducing themselves to the group and answering the question ``How would you describe what makes dance meaningful to you?'' to consider aspects of dance they wanted to preserve in their later design exercise. We then shared survey themes around the benefits and challenges for accessible \revised{asynchronous} movement education. 

We then presented a ``toolbox'' of non-visual teaching modalities with discussions and examples. Participants first discussed \textit{``What are non-visual ways to teach dance?''} in pairs, then shared with the larger group (i.e. think-pair-share~\cite{kaddoura2013think, apriyanti2020think}). We then presented examples of verbal audio description for dance, then examples of onomatopoeia and movement sonification (where movement qualities create sound). Following the examples, we asked participants via think-pair-share to discuss what was helpful or unhelpful in the examples for conveying movement. 
We then presented and discussed haptics as a medium for movement instruction, explaining existing applications such as virtual exergames~\cite{morelli2010vibowling} and precise hand navigation~\cite{hong2017wristhaptics}. \\

\noindent \textit{\textbf{Teaching \& Designing: 85 minutes. }}Following the introductory presentation, we split participants into groups of 3-5 people with at least one BLV participant and one dance instructor per group (Table~\ref{tab:groups}). We assigned each group a dance tutorial clip around 2.5 minutes long as a reference point for creating concrete designs of dance tutorial systems. We typically formed groups such that multiple participants shared experience in a common dance style, and we selected the tutorial clip to match the dance style experience of the group. \revised{The dance styles in the provided clips included Salsa, Contemporary, Hip-hop, Pop, and Cha-cha. As movement structures and teaching methods vary across different dance styles~\cite{conroy2025stanfordphilosophyofdance}, the diversity of dance styles in our sample clips allowed for a wider range of teaching techniques across groups. 
}
We also ensured that selected clips were narrated and suitable for beginners. 
Partcipants remained with only their groups for the \textbf{teaching} and \textbf{designing} stages. 

 We asked dance instructors to \textbf{teach} BLV participants the content from the given dance tutorial clip to establish common ground and explore concrete examples of non-visual instruction. 
We told dance instructors to pay attention to how they narrate instructions, provide feedback on movements, use tactile methods, and/or use sound cues and onomatopoeia. We told learners to identify techniques that were helpful or unhelpful.


\begin{table}[h!]
\centering
\renewcommand{\arraystretch}{1.3} %
\begin{tabular}{l l l p{2cm}}  
\toprule
\textbf{Workshop} & \textbf{Group} & \textbf{Participants} & \textbf{Dance Style} \\
\midrule
1 & G1 & B1, D1, H1 & Salsa \\
1 & G2 & B2a, B2b, D2, S2 & Contemporary \\
1 & G3 & B3a, B3b, D3 & Hip-hop \\
2 & G4 & B4, D4a, D4b & Contemporary \\
2 & G5 & B5a, B5b, D5 & Contemporary \\
2 & G6 & B6, D6, S6 & Salsa \\
3 & G7 & B7a, B7b, D7a, D7b & Pop \\
3 & G8 & B8a, B8b, D8a, D8b, A8 & Cha-cha \\
\bottomrule
\end{tabular}
\caption{Participant allocation across different groups and the dance style taught in the tutorial clip given to each group.}
\label{tab:groups}
\end{table}

\begin{figure}[t]
    \centering
    \includegraphics[width=3.33in]{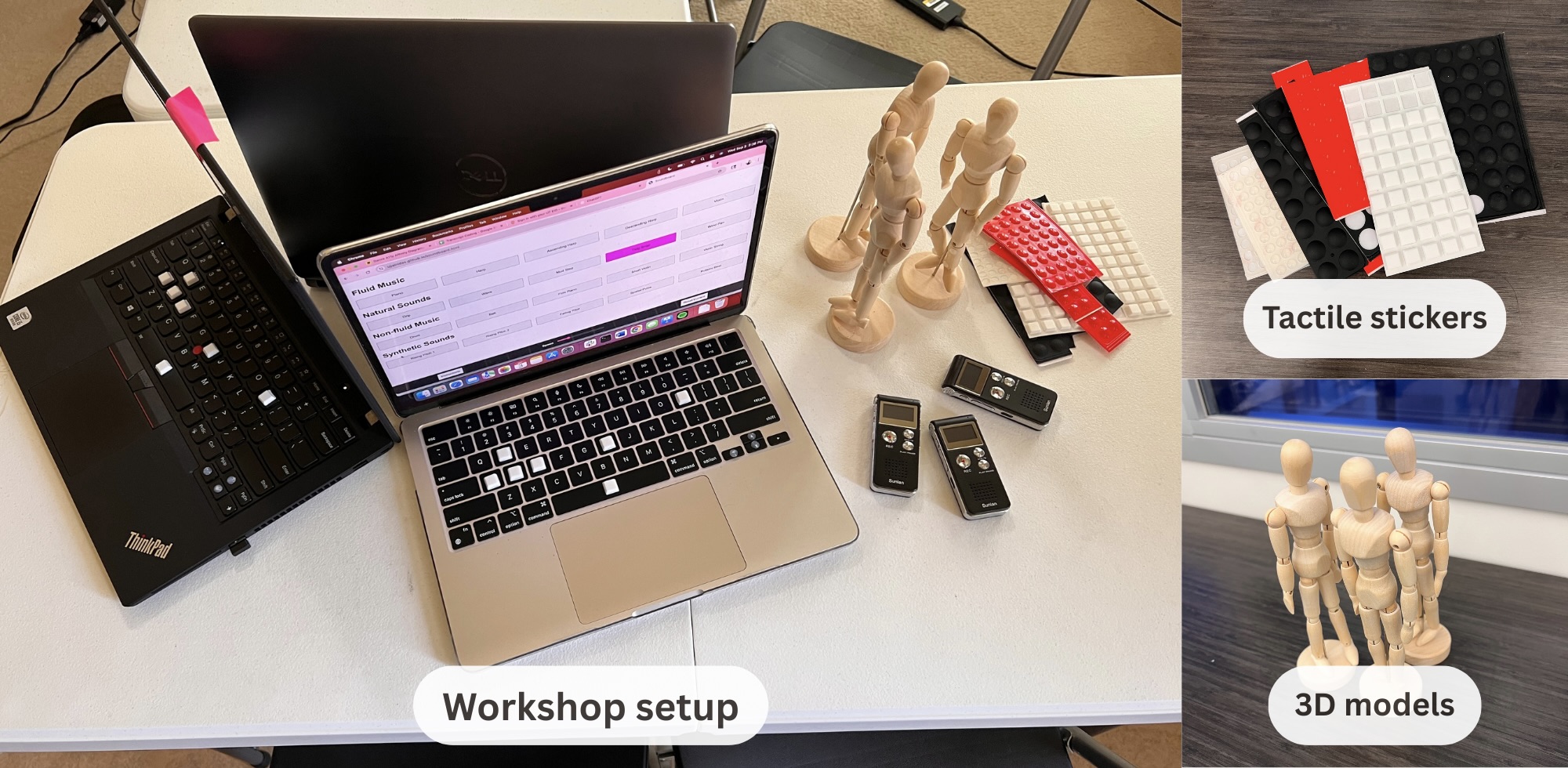}
    \caption{We provided participants with an accessible soundboard application and voice recorders, configurable 3D models, and tactile stickers that participants used to embody haptic device placements.}
    \label{fig:tools}
\end{figure}
\begin{figure}[t]
    \centering
    \includegraphics[width=3.33in]{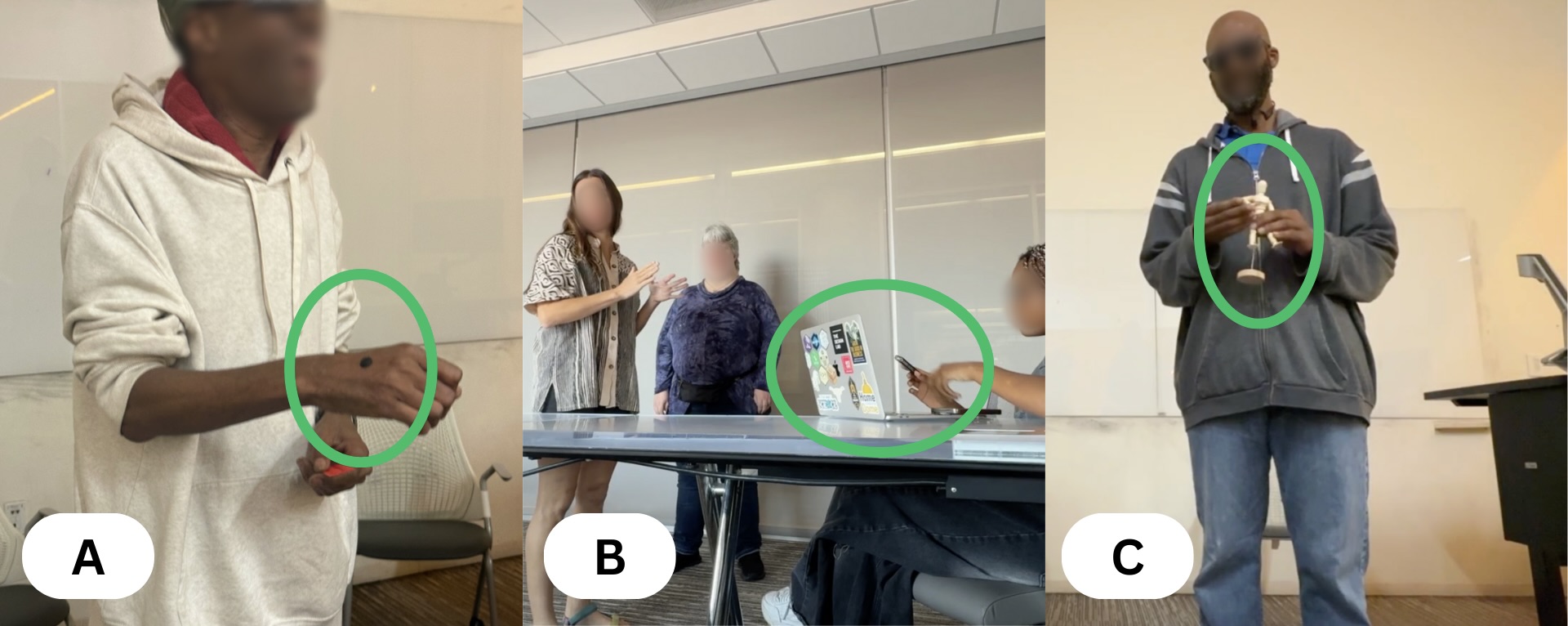}
    \caption{Participants using tactile stickers to mimic haptic device placement (A), voice recorders to record verbal descriptions and sound cues from the soundboard (B), and the 3D models to learn poses (C).}
    \label{fig:peopleusingtools}
\end{figure}

Within their small groups, participants then discussed and \textbf{designed} \textit{a non-visual tutorial system for \revised{}{asynchronous} dance education}. \revised{We also asked participants to create \textit{a concrete prototype or a walkthrough} of an accessible asynchronous dance instruction system using their given clip as an example.} We provided participants with several \textbf{prototyping materials} to facilitate the designing stage (Figure~\ref{fig:tools}): \textit{recording devices} with raised buttons for recording verbal narration and non-verbal sound cues; a custom accessible \textit{soundboard} application that allowed preview and adjustment of fluid/non-fluid music, natural sounds, and synthetic sounds (as informed by movement sonification literature~\cite{frid2018motionfluidity, landry2020interactive}); \textit{3D movable human figures} for demonstrating poses and haptic device placements; \textit{tactile stickers} for prototyping haptic device placement on provided 3D figures or on participants themselves. Figure~\ref{fig:peopleusingtools} shows participants using these prototyping tools to brainstorm and design their systems. \\

\begin{figure}
    \centering
    \includegraphics[width=3.33in]{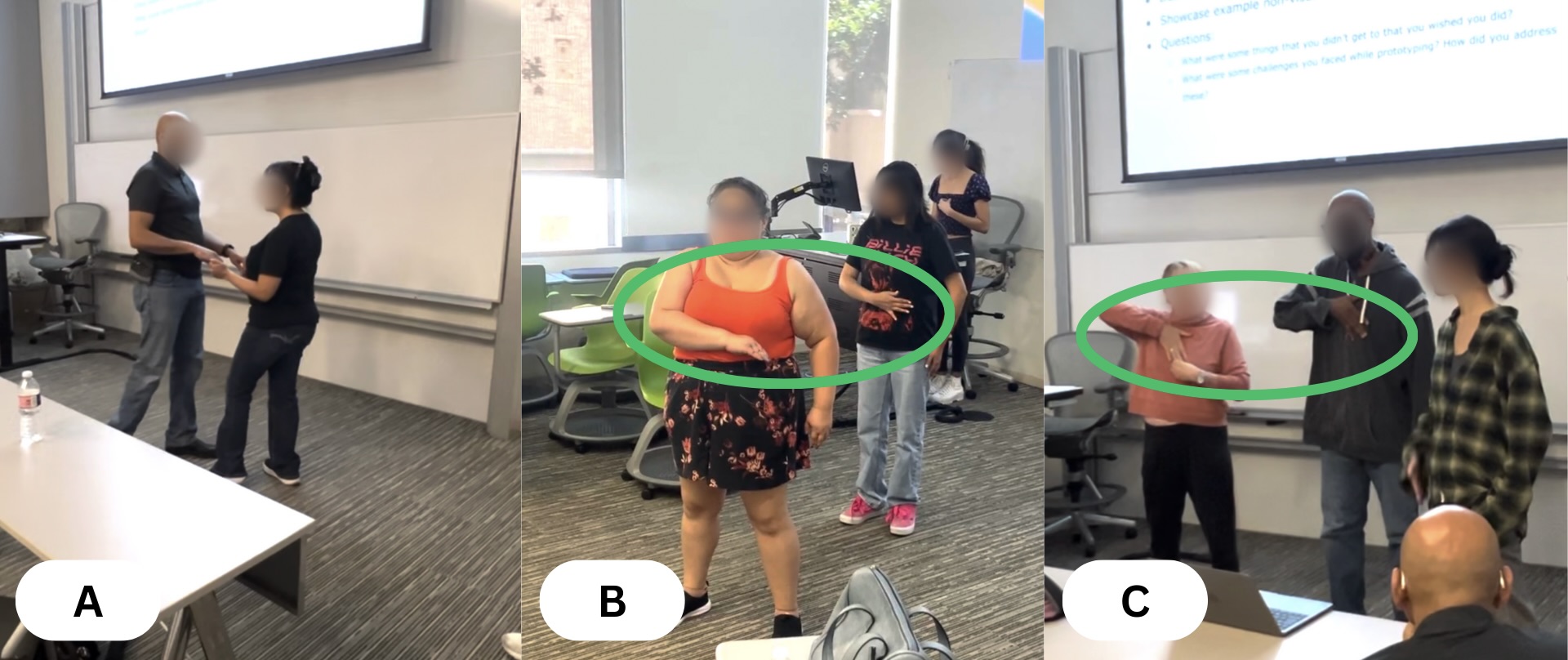}
    \caption{Participants presenting their tutorial systems during the last stage of the workshop by physically demonstrating haptic cues (A) and verbal narration and sounds (B and C).}
    \label{reenactments}
\end{figure}

\noindent \textit{\textbf{Presenting: 30 minutes.}}
Following group design sessions, all groups reconvened and were each asked to explain their non-visual tutorial system workflow in-depth, shown in Figure~\ref{reenactments}. We iterated on the workshop design following Workshop 1 and asked participants to additionally physically \textbf{re-enact} their system, using their given dance tutorial clip as an example. Each group designated one or more participants to act as the \textit{system}, providing instruction, haptic and tactile feedback through (with permission) physical touch, and sound cues through the soundboard and onomatopoeia. Groups designated one or more participants to act as the \textit{learners} from the system. Participants narrated system interactions during these re-enactments.

\subsubsection{\textbf{Synchronous, In-Person Workshops for Asynchronous, Remote Dance Instruction Systems}}
\revised{Although our workshops addressed asynchronous dance instruction, we held them in-person to allow for improved collaboration between participants through a shared physical space. The in-person setting allowed participants to establish common ground when learning and teaching one another example dances. Additionally, participants were able to use hands-on prototyping materials when designing their systems. During prototyping and presentations, participants limited system interactions to what would be available in asynchronous contexts. For example, the use of touch was limited to where haptic devices would be placed, or verbal descriptions were limited to what would be provided by the system. Each group was accompanied by a member of the research team to center discussion on asynchronous contexts. Our in-person setup therefore allowed for engaging, collaborative workshops while addressing our problem space of asynchronous dance instruction.}

\subsection{Data Collection and Analysis}
All workshops and group sessions were audio-video recorded. Survey results, transcripts, and videos from session recordings were coded using existing frameworks for qualitative thematic analysis~\cite{braun2006using, khandkar2009open}. After thorough open-coding of transcripts and survey results by four members of the research team, resulting codes were collaboratively organized into themes addressing RQ1 and RQ2. 
\section{Results}

In this section, we share participants' strategies and challenges of current dance instruction, their preferred approaches for using different modalities in remote instruction, and their concepts for future multimodal systems for end-to-end \revised{asynchronous} dance instruction.

\subsection{Current Strategies and Challenges}
To provide context for workshop solutions, we briefly summarize\footnote{As future systems are the primary focus of our study rather than existing practices, we include the extended analysis of existing practices in the Supplementary Materials.} participants' existing practices for accessible dance instruction, and the challenges they experienced watching the existing dance videos in the teaching phase. 

BLV learners with dance experience and dance instructors with experience instructing BLV learners all reported that their prior experience was with \textit{synchronous, in-person} instruction. The dominant mediums for instruction were \textbf{verbal instructions} and \textbf{touch}. For in-person instruction, B2a described teaching a class of BLV children, saying, \textit{``I would be at the front so they could follow me and yell what we were doing at the same time.''} B3b, who had experience with dance choreography during saxophone performances, described in his survey response: \textit{``My sighted section leader would stand behind to me and put his hand on the center of my back and he would tap patterns with his fingers. If he tapped my right shoulder blade that would mean step right. Or step back in.''} All groups also emphasized the importance of \textbf{feedback} in existing dance instruction methods. G4 discussed how providing feedback is easier with synchronous instruction, as when watching dance videos learners can not ask questions (\textit{e.g.}, \textit{``Am I doing this right?''}). 
G2, G4, and G8 noted that current synchronous dance teaching practices allow instructors to \textbf{personalize} their instructions to students' individual needs and progress. 

All groups watched the assigned example dance clips during the teaching phase to learn the sequence of movements, and all groups identified major challenges with directly following the videos alone: \textbf{lack of context}, \textbf{lack of detail}, and \textbf{vague terminology}. For example, G8's clip directly began demonstrating the cha-cha routine without giving context such as how many steps and movements there were. All clips did not describe orientations of body parts with enough detail to recreate the movements. 
B8a expressed: \textit{``It's not clear what step she is doing or if she is stepping at all. She could be stepping, or just shifting weight.''}

\subsection{Modalities for Dance Instruction}

Participants used diverse strategies to convey motion in the small group sessions (Teaching phase, Designing phase): verbal descriptions, non-verbal audio, and haptic or tactile approaches. 



\subsubsection{\textbf{Verbal Descriptions}}
All groups emphasized the importance of clear, consistent, and well-paced language that could describe movement mechanics, orientation, expression, and timing. These descriptions were broken down into manageable pieces, outlined into distinct vocabularies, incorporated with metaphors, and adapted based on learner experience and preferences. A8 explained the importance of descriptiveness in helping BLV learners visualize movements:
\begin{quote}
   \textit{ ``When you're teaching somebody, you have to talk every minute, every step. Everything that you're doing, you have to talk about it and say what you're doing. Because that's the only way a person is going to be able to do it.''}
\end{quote}

\paragraph{\textbf{High-level context}}
G5, G6, G8 provided high-level, \\cross-movement context, either before or during instruction of the routine. For example, before teaching, A8 explained the cha-cha clip by saying, \textit{``The dance is working in opposites [...] so if you go forward, you will complement that by going backwards.''} B8a noted that descriptions should first explain the rhythm using verbal counts before moving into defining the steps. On the other hand, B5a and B5b described the rhythm after movement definitions. D6 noted how it was better to use semantic cues that reference movements that happen at a specific rhythm, such as \textit{`forward'}, \textit{`back'}, and \textit{`close'}, instead of counts.

\aptLtoX{\begin{shaded}
\noindent \textbf{Takeaway:} Provide high-level, cross-movement context on dance routine patterns and rhythm. 
\end{shaded}}{
\begin{center}
\setlength{\fboxrule}{0.8pt} 
\setlength{\fboxsep}{4pt}    

\fcolorbox{black}{gray!10}{%
  \parbox{\dimexpr\linewidth-2\fboxsep-2\fboxrule\relax}{%
    \hspace*{2pt}
    \textbf{Takeaway:} Provide high-level, cross-movement context on dance routine patterns and rhythm. 
    \hspace*{2pt}
  }%
}
\end{center}}

\paragraph{\textbf{Breaking down dance routines}}
Dance instruction involves teaching \textit{routines}, or a sequence of movements occurring one after another. Some dances have designated names for specific movements that can facilitate the instruction of movements. However, D6, a dance historian, explained that current terms are not consistent across different dance styles or routines: \textit{``If I say the word} forward\textit{, that could be a whole bunch of different things. [...] That could mean five steps forward or one step forward.''} 

All groups approached teaching routines verbally by breaking down routines into distinct movements, and labeling each movement of a routine in a \textbf{movement vocabulary}, examples of which are shown in Table \ref{tab:movementvocab}. B7a and B8b compared this vocabulary to a \textit{'table of contents'} and a \textit{'set of line items for the dance'}, respectively.

G3, G4, G5, G6, G7, and G8 emphasized the importance of clearly defining the \textbf{start position} of a dance routine, including the starting positions as an item in their movement vocabularies such as shown in Table \ref{tab:movementvocab}. B5a explained: \textit{``Many people who try to learn from the video can see what she's doing, but it doesn't work for us. So trying to establish what's my neutral position. Where my feet are, where are my arms?''}

G2, G6, and G8 used memorable \textbf{movement names} that summarized the movement, such as \textit{`Beach Ball'}, \textit{`Water Scoop'}, and \textit{`Close'}. G8's vocabulary terms were pre-existing cha-cha step names, such as \textit{`Rock Step'} and \textit{`Triple Step'}, that were re-defined in the context of teaching BLV students. G1, G3, G4, and G5 labeled movements \textbf{numerically}, such as \textit{`Movement 1'} or \textit{`Count 3'}. A drawback of a numerical approach is that numbers are less intuitive than names, and would require learners to practice the movements more extensively to internalize their definitions. G6 additionally mentioned that numbers can sometimes indicate timing, so numbers mapped to movements with different durations can be confusing. 

Once the movement vocabulary is defined, the terms in these vocabularies are used as a shorthand to \textit{reference} rather than \textit{re-describe} the movement. Examples said by participants are: \textit{``Make a Beach Ball to your right''}, \textit{``Rock Step forward, Rock Step back, right foot back to center''}, and \textit{``let's do Movement 5 a few times in a row''}.

\aptLtoX{\begin{shaded}
\noindent  \textbf{Takeaway:} Break down dance routines into a distinct set of defined terms, or \textit{movement vocabularies}. Use these terms as shorthand to reference movements during instruction.
\end{shaded}}{
\begin{center}
\setlength{\fboxrule}{0.8pt} 
\setlength{\fboxsep}{4pt}    

\fcolorbox{black}{gray!10}{%
  \parbox{\dimexpr\linewidth-2\fboxsep-2\fboxrule\relax}{%
    \hspace*{2pt}
    \textbf{Takeaway:} Break down dance routines into a distinct set of defined terms, or \textit{movement vocabularies}. Use these terms as shorthand to reference movements during instruction.%
    \hspace*{2pt}
  }%
}
\end{center}}

\begin{table*}[h!]
\centering
\renewcommand{\arraystretch}{1.4}
\resizebox{\linewidth}{!}{
\begin{tabular}{l l p{8cm} l}
\toprule
\textbf{Group} & \textbf{Movement Name} & \textbf{Definition} & \textbf{Technique} \\
\midrule
G2 & Water Scoop & Take your right arm down. Go from the bottom right and brush your pinky on your thigh as you go all the way to the top left across your body like you’re scooping water, and bring it back down. & Body parts in relation \\
G3 & Count 0 & This is your starting position where you are facing forward with your legs about hip width apart. & Incorporating starting position \\
G4 & Movement 5 & Straighten your right leg, slide it out to the right, and shift to your weight to your left foot as you stand on your right toes. At the same time, lift your right arm over your head and lean to the left. & Overall body description \\
G5 & Count 1 & The end goal is you and your body is going to twist to the right. Your right foot is going to point outward towards the right. With your hand, kind of like a sword you’re going to make a slicing motion across your body, ultimately pointing in the same direction that your foot is pointing at. & Global to local description \\
G8 & Triple Step & \textit{[Instructor puts hands on learner's back]} These are my feet. Your weight is here \textit{[add pressure to left hand]} and then this is where your toe is \textit{[remove pressure from right hand]}. So you're going to step out to the right \textit{[perform step on back]} 1, 2, 3. & Multimodal vocabulary \\
\bottomrule
\end{tabular}}
\caption{Example movement vocabulary terms and corresponding techniques used in definitions.}
\label{tab:movementvocab}
\end{table*}

\paragraph{\textbf{Breaking down individual movements}}
Participants expressed that extensive, detailed verbal descriptions can be overwhelming for listeners and detract from the fun of learning dance. When defining the terms in their movement vocabularies, G4, G5, G6, and G7 concluded that definitions should be as descriptive as possible, but broken down into further pieces that are more manageable for learners. These pieces included descriptions of overall body movements and positioning, individual body part movements, and the movement of different body parts in relation to one another. 

Many dance tutorials involve multiple body parts moving at the same time, as D4a emphasized: \textit{``Dances will rarely be in just one body part. It will be in your leg, arm, torso, head, and other places at the same time.''} All groups found it important to \textbf{local descriptions} on what each body part is doing, including parts that remain still. D3 added on that it's important to include how to orient individual body parts during motions, and ``describing where each body part is on the 360 degree wheel.'' Similarly, G2 and G5 discussed individual orientations of specific body parts such as the palms and arms. 

While creating movement descriptions, groups created their own terms. D3 said `\textit{`plant''} and \textit{``stagnant''} while describing the motions from the clip to B3a and B3b. She explained the terms separately, stating:

\begin{quote}
\textit{``When I'm saying \textit{plant}, that means in conjunction to whatever body part is moving. Like whenever I say pick something up and then \textit{plant} it, it is usually referred to the same body part. [...] And then if I'm saying something is \textit{stagnant} [...] it's just still and [...] we're not engaging it.''}
\end{quote}

G2, G5, G6, and G8 found that describing \textbf{how body parts moved in relation} to one another provided additional grounding for learners. For example, D2 explained arm movements relative to the body for the \textit{`Water Scoop'} in Table \ref{tab:movementvocab}. G5 discussed how it is helpful to describe how individual body movements effect each other. For example, B5a explained, \textit{``If you bend your knees, that will move your hips.''} G6 and G8 described distance relative to other body parts. For example, B8a explained, \textit{``You're gonna take a sidestep to your right. It’s about the width of your shoulders, maybe a but less.''}

G2, G4, G5, and G6 prioritized describing \textbf{overall body positioning} and weight distribution. For example, D2 provided descriptions such as \textit{``Once you hit the Beach Ball, everything should come into the center of your body''} and D4 explained weight shifts in \textit{`Movement 5'} in Table \ref{tab:movementvocab}.

Groups had two strategies to describe movements: either local to global (G2, G3, G4) or global to local (G5). For example, G3 started at the legs before moving up the body to the head, describing body movements individually before having participants practice it as one motion. G5, on the other hand, explained changes in overall body positioning before describing individual body parts.

\aptLtoX{\begin{shaded}
\noindent   \textbf{Takeaway:} Break down individual movements by locally describing body parts, body parts in relation to one another, and overall body positioning.
\end{shaded}}{
\begin{center}
\setlength{\fboxrule}{0.8pt} 
\setlength{\fboxsep}{4pt}    

\fcolorbox{black}{gray!10}{%
  \parbox{\dimexpr\linewidth-2\fboxsep-2\fboxrule\relax}{%
    \hspace*{2pt}
    \textbf{Takeaway:} Break down individual movements by locally describing body parts, body parts in relation to one another, and overall body positioning.%
    \hspace*{2pt}
  }%
}
\end{center}}

\paragraph{\textbf{Verbal feedback}}

G3, G5, G6, and G8 incorporated back-and-forth \textbf{incremental verbal feedback} into their system designs, along with \textbf{positive affirmations} if movements were done correctly. For example, G8's system allowed users to ask clarifying questions, such as \textit{``Does this position look good?''}, with incremental feedback, such as \textit{``No, you want to move your foot a little to the left.''} 

G6 discussed have a system \textbf{adapt verbal descriptions} given knowledge of the learner's movements and understanding of the descriptions. To illustrate this concept, D6 first effectively described movements to B6 by watching her and assessing her understanding of the instruction. However, D6 then turned around and attempted to teach B6 a movement, which proved to be difficult because of his inability to gauge B6's correctness and understanding. G5's system included prompts between learning steps that ask if the user feels comfortable performing the movement. If not, additional instruction is provided or the given instructions are repeated. An example prompt was, \textit{``In the neutral position, do you feel comfortable?''}. 

\aptLtoX{\begin{shaded}
\noindent  \textbf{Takeaway:} Provide verbal feedback by offering incremental corrections. Adapt verbal descriptions based on the learner's correctness and understanding of current descriptions.
\end{shaded}}{
\begin{center}
\setlength{\fboxrule}{0.8pt} 
\setlength{\fboxsep}{4pt}    

\fcolorbox{black}{gray!10}{%
  \parbox{\dimexpr\linewidth-2\fboxsep-2\fboxrule\relax}{%
    \hspace*{2pt}
    \textbf{Takeaway:} Provide verbal feedback by offering incremental corrections. Adapt verbal descriptions based on the learner's correctness and understanding of current descriptions. %
    \hspace*{2pt}
  }%
}
\end{center}}

\paragraph{\textbf{Overall verbal strategies}}
Learner familiarity with the dance routine or style determines whether to use full descriptions or movement vocabulary terms as a shorthand. G2, G4, G5, G6, G7, and G8 \textbf{varied the descriptiveness} of verbal instructions, either as the learner got more advanced or based on their level of experience. For example, G4 noted that it would be efficient to convert to shorter phrases once learners have understood the motions, such as: \textit{`arms to side'} and \textit{`bend knees'}. On the other hand, B8a was already familiar with a \textit{`Triple Step'} in cha-cha and did not need that movement defined. 

G5 and G8 found it helpful to \textbf{repeat words and phrases} as students learned and practiced movements. B6, for example, expressed how repetition helps students remember steps and correct mistakes: \textit{``I personally like [redundancy]. The more you say it, the more it sticks. Keep saying `left', keep saying `right'''} 

All groups emphasized using \textbf{metaphors} that are relatable to BLV learners. These were useful in conveying structure, emotions, and fluidity associated with movements. Groups used a combination of \textit{tactile} (related to touch and sensations), \textit{kinesthetic} (related to actions), and \textit{visual} metaphors (related to how specific objects look or move). Tactile metaphors were most common. Table \ref{tab:metaphors} highlights example metaphors used by groups in their verbal descriptions. 

\aptLtoX{\begin{shaded}
\noindent \textbf{Takeaway:} Vary descriptiveness based on learner's familiarity with the dance style and routine. Use repetition of words and phrases to improve memory of movements. Use metaphors to convey temporal and expressive qualities.
\end{shaded}}{
\begin{center}
\setlength{\fboxrule}{0.8pt} 
\setlength{\fboxsep}{4pt}    

\fcolorbox{black}{gray!10}{%
  \parbox{\dimexpr\linewidth-2\fboxsep-2\fboxrule\relax}{%
    \hspace*{2pt}
    \textbf{Takeaway:} Vary descriptiveness based on learner's familiarity with the dance style and routine. Use repetition of words and phrases to improve memory of movements. Use metaphors to convey temporal and expressive qualities. %
    \hspace*{2pt}
  }%
}
\end{center}}

\begin{table}[h!]
\centering
\begin{tabular}{l l p{4.5cm}}  
\toprule
\textbf{Group} & \textbf{Metaphor Type} & \textbf{Metaphor} \\
\midrule
G2 & Tactile & Move as if you are scooping water with your hand. \\
& Tactile & Hold a bubble to your side. \\
& Tactile & Let that arm pull you to the right. \\
 & Kinesthetic & Reach your hands out in front of you as if you are holding a beach ball. \\
G3 & Tactile & Move your arm through honey. \\
G4 & Tactile & Your arms are floating in water. \\
 & Kinesthetic & Your arms are in front of you in first position, like you’re hugging a tree \\
 G5 & Tactile & Your hand is a sword slicing through the air. \\
 & Visual & Your hand will move like a bike wheel that will come towards you. \\
G6 & Kinesthetic & Imagine you’re in a really narrow hallway, and if you put your elbows out to the side, your elbows would hit the walls. Imagine you have crayons on the tips of your elbows and you’re just drawing little scribbles on the walls. \\
G8 & Visual & Like a mermaid, your ankles are attached \\
\bottomrule
\end{tabular}
\caption{Example metaphors used to define movement names and corresponding metaphor types.}
\label{tab:metaphors}
\end{table}

\subsubsection{\textbf{Non-Verbal Audio}}
Participants used sound cues and music to indicate specific movements, tempo, pacing and transitions, and feedback. Groups discussed considerations for combining sound with existing music. 

\paragraph{\textbf{Sound-to-movement mappings}}
Six groups found non-verbal audio helpful in creating sound-to-movement mappings, where each movement from the movement vocabularies are mapped to a distinct sound cue. G2 and G7 noted that this association would improve memory of the movements and provide independence for learners. Sound mappings were both direct (based on matching motion qualities to sound qualities) and semantic (based on metaphorical meanings of movements that match certain sounds). 

G2, G4, and G5 incorporated \textbf{semantic mappings}. G2 chose sounds that metaphorically matched the how movements feel, such as a water splash sound to convey the fluidity of an arm movement. Additionally, G4 used a wind blowing sound for a three-step turn, which participants felt was reminiscent of a person being blown in the wind. D5 noted that contemporary dance allows for more freedom in how dancers decide to move. As such, their system provided different options for textures that learners could exhibit, each with a different sound cue. 

G2, G3, G4, G5, G7, and G8 used \textbf{direct mappings}. G2, G3, G4, and G7 chose sounds matching the speed and rhythm of movements. For example, G2 used a wind sound that had the same speed and abrupt stop as the movement it was mapped to. G4 mapped the slow raising of the arms to the side of the body to a soft \textit{violin vibrato} sound of the same duration. On the other hand, the abrupt moving of the hands to the chest was mapped to a \textit{quick drum} sound. G4 additionally matched pitch to height for some sound cues — for example, they used a \textit{rising harp} sound to indicate the arms moving up, and a \textit{descending harp} sound to indicate a lunge. G8 decided to have a cue for each step or weight shift to match the rhythm. However, they differentiated between different types of steps with the sound being played. For example, a \textit{cha} sound was used for each part of a Triple Step, whereas a \textit{tha} sound was used for each rock in a Rock Step. To indicate directions, G8 incorporated spatial audio into sound-to-movement mappings. For example, \textit{``Triple Step''} heard from the right ear would signal to the user that they should do a Triple Step to the right. 

\aptLtoX{\begin{shaded}
\noindent \textbf{Takeaway:} Map items in the movement vocabulary to distinct sound cue, either semantically through metaphorical sounds or with direct associations between physical and sound attribute.
\end{shaded}}{
\begin{center}
\setlength{\fboxrule}{0.8pt} 
\setlength{\fboxsep}{4pt}    

\fcolorbox{black}{gray!10}{%
  \parbox{\dimexpr\linewidth-2\fboxsep-2\fboxrule\relax}{%
    \hspace*{2pt}
    \textbf{Takeaway:} Map items in the movement vocabulary to distinct sound cue, either semantically through metaphorical sounds or with direct associations between physical and sound attribute. %
    \hspace*{2pt}
  }%
}
\end{center}}


\paragraph{\textbf{Sound for rhythm, pacing, and transitions.}}
G1, G2, G4, G7, and G8 used sound to ground learners in a dance routine. For example, G1 used a rising pitch to indicate the start of the routine and a falling pitch to indicate the end of a routine. Similarly, G7 used a bell sound to indicate the start of a routine. B7a emphasized how sound cues could help learners establish \textit{when} they should be doing certain movements: \textit{''If you hear that sound, and you're not squatting yet, you know that you're behind.''} G8 described how a distinct sound could play to indicate the transition \textit{between} steps, such as switching directions from left to right.

G2, G5, and G7 used sound cues to help maintain \textbf{rhythm}. G7 used tapping and clapping sounds in their instruction to establish rhythm. On the other hand, G2, G3, and G5 used a metronome sound in the background of their tutorial to establish the tempo of the dance. 

\aptLtoX{\begin{shaded}
\noindent \textbf{Takeaway:} Use sounds cues to ground learners on when movements happen, transitions within a routine, and rhythm of the routine.
\end{shaded}}{
\begin{center}
\setlength{\fboxrule}{0.8pt} 
\setlength{\fboxsep}{4pt}    

\fcolorbox{black}{gray!10}{%
  \parbox{\dimexpr\linewidth-2\fboxsep-2\fboxrule\relax}{%
    \hspace*{2pt}
    \textbf{Takeaway:} Use sounds cues to ground learners on when movements happen, transitions within a routine, and rhythm of the routine. %
    \hspace*{2pt}
  }%
}
\end{center}}

\paragraph{\textbf{Sound for emotions.}}
G5 used background tracks \textit{separate from the music of the dance}, called \textit{`Incidental music'}, to establish the energy and emotion in their routine. D5 noted: \textit{``In contemporary dance, we will have music softly playing in the background that isn’t giving a strict idea of the tempo, but is giving some idea of how the movement should be done.''} He then played an example sound track and asked participants to describe where the music took them emotionally before beginning instruction.

\aptLtoX{\begin{shaded}
\noindent \textbf{Takeaway:} Incidental music playing in the background during instruction may help establish emotion and energy of a dance.
\end{shaded}}{
\begin{center}
\setlength{\fboxrule}{0.8pt} 
\setlength{\fboxsep}{4pt}    

\fcolorbox{black}{gray!10}{%
  \parbox{\dimexpr\linewidth-2\fboxsep-2\fboxrule\relax}{%
    \hspace*{2pt}
    \textbf{Takeaway:} Incidental music playing in the background during instruction may help establish emotion and energy of a dance. %
    \hspace*{2pt}
  }%
}
\end{center}}

\paragraph{\textbf{Sound cues with existing music.}}
G1, G2, G6, and G7 discussed ways to combine sound cues with existing dance music. G2 and G6 noted that these combined sounds may be confusing. However, D6 highlighted that music is typically not included in instruction until after the movements are learned. Thus, these groups envisioned systems where sound cues would be introduced to aid instruction on movements before incorporating the music of the dance. 

Furthermore, B1 preferred overlaying sound cues with music rather than words, as they were easier to process and pinpoint during instruction. Similarly, G7 discussed how overlaying sound cues can be used to show learners when movements take place once they have learned the movements themselves. G1, G2, and G4 used elements such as volume, pitch, and timbre to distinguish sound cues from the existing music.

\aptLtoX{\begin{shaded}
\noindent \textbf{Takeaway:} Introduce sound cues before incorporating dance music into instruction. When overlaying sound cues on music, use distinguishing elements, such as volume, pitch, or timbre. 
\end{shaded}}{
\begin{center}
\setlength{\fboxrule}{0.8pt} 
\setlength{\fboxsep}{4pt}    

\fcolorbox{black}{gray!10}{%
  \parbox{\dimexpr\linewidth-2\fboxsep-2\fboxrule\relax}{%
    \hspace*{2pt}
    \textbf{Takeaway:} Introduce sound cues before incorporating dance music into instruction. When overlaying sound cues on music, use distinguishing elements, such as volume, pitch, or timbre. %
    \hspace*{2pt}
  }%
}
\end{center}}

\paragraph{\textbf{Sound as feedback}}

G3, G6, and G8 incorporated sound cues as feedback. After deliberating on whether to give sound cues for both correct and incorrect motions, G6 decided that sound cues should play only for \textit{positive feedback}. This would reduce discouragement and be less overwhelming for learners, as learners would hear an occassional positive cue rather than a constant series of negative cues as they attempt the movement multiple times.

G1, G3, and G6 incorporated live sound feedback on learners' movements. G1 envisioned spatial audio providing \textit{live feedback} on a learners movements. For example, as they moved their arm to the right, the sound feedback would pan to the right. B3b outlined a system where height and radius of the arm would be mapped to pitch and vibration intensity, respectively. G6 noted that having a sound of varying pitch to indicate how far a body part is from the target may be useful for conveying exact hand positioning. However, they concluded that \textit{simpler sonification algorithms} with fewer variable elements would be easier for students to understand. For example, G6 discussed that one-dimensional sound feedback (such as only pitch mapped to height) is better than two or three-dimensional feedback. 

\aptLtoX{\begin{shaded}
\noindent \textbf{Takeaway:} Use sound cues to deliver \textit{positive feedback} on movements. When providing live feedback, use simple sonification mappings to convey body part positioning.
\end{shaded}}{
\begin{center}
\setlength{\fboxrule}{0.8pt} 
\setlength{\fboxsep}{4pt}    

\fcolorbox{black}{gray!10}{%
  \parbox{\dimexpr\linewidth-2\fboxsep-2\fboxrule\relax}{%
    \hspace*{2pt}
    \textbf{Takeaway:} Use sound cues to deliver \textit{positive feedback} on movements. When providing live feedback, use simple sonification mappings to convey body part positioning. %
    \hspace*{2pt}
  }%
}
\end{center}}

\subsubsection{\textbf{Haptic and Tactile Input}} 
G1, G3, G4, G6, G7, and G8 outlined the use of wearable haptic devices, tactile floor markers and mats, and 3D models for instruction and feedback. G7 noted how it is important for BLV learners that have haptic devices going on different places of the body to be easily differentiable through size, shape, or texture. 

\paragraph{\textbf{On-body instruction.}}

G1, G3, G4, G7, and G8 explored the use of on-body haptics for movement instruction. G1 proposed having haptic bands around the learner's ankles or knees to instruct \textbf{limb placement}. Ankle devices would vibrate in the direction of movement, with intensity of vibration increasing with the end position's distance from the center. D1 compared this to placing an elastic band around the feet, where tension should increase the further your feet move from the starting point. 

G8's cha-cha system involved a haptic device on each of the learner's hands and on the back. Back-worn haptics would be used to first show feet placements for each of the step types. This was reenacted with B8a putting her hands on B8b's back, shown as \textit{`Triple Step'} in Table \ref{tab:movementvocab} Hand-worn haptic devices were used to indicate left and right directions. With both back-worn and hand-worn devices, the intensity of the vibration would indicate the weight distribution of the step. D8a explained, \textit{``If you're putting all of your weight on one foot, maybe the intensity could be bigger. And then a slightly lighter, softer vibration means a tap with the toe, like with the Bachata.''} 

G7 proposed haptic devices on specific moving body parts. For their dance clip, this was the right arm and the left knee. Both devices would vibrate at the same time and keep vibrating for the duration of the movement to convey speed and length. Similarly, G8 discussed having hand-worn haptic devices provide consistent vibrations for the duration of a step to differentiate between steps of different lengths. 

\aptLtoX{\begin{shaded}
\noindent  \textbf{Takeaway:} Haptics for asynchronous learning can function as a parallel to touch feedback to instruct foot positioning, body weight distribution, and movement duration.
\end{shaded}}{
\begin{center}
\setlength{\fboxrule}{0.8pt} 
\setlength{\fboxsep}{4pt}    

\fcolorbox{black}{gray!10}{%
  \parbox{\dimexpr\linewidth-2\fboxsep-2\fboxrule\relax}{%
    \hspace*{2pt}
    \textbf{Takeaway:} Haptics for asynchronous learning can function as a parallel to touch feedback to instruct foot positioning, body weight distribution, and movement duration. %
    \hspace*{2pt}
  }%
}
\end{center}}

\paragraph{\textbf{On-body feedback}}
Haptics served as both confirmation of correct \textbf{positioning} and \textbf{negative feedback} to indicate mistakes. B3a mentioned that she thought about haptics being useful in the `reverse': \textit{``I would want haptics to tell me not when to move, but when to \textit{stop} moving. For example, if I've moved too far, it would vibrate.''}

G1's prototype included devices around the hips and on the lower back. When a dancer had to rotate, the vibration would start in the direction that you are facing and would move to the direction of the rotation. H1 compared this system to a compass, where the direction of vibration always indicates the direction that the dancer should face. 

G4 and G6 discussed that for dances where hand positions matter, having haptics or sound indicate when you've hit the correct position would be helpful. G3 discussed a system where there would be haptic bracelets on each wrist and ankle providing live feedback on hand positioning by vibrating with more intensity as the hands moved away from the starting position. However, S6 mentioned his experience of not being able to feel on-body haptic devices while moving during live performances. This outlines the importance of calibrating haptic device intensity, as discussed by G8, to be at a comfortable and helpful level. 

\aptLtoX{\begin{shaded}
\noindent \textbf{Takeaway:} Wearable haptics can be used to provide feedback on limb positioning and body orientation.
\end{shaded}}{
\begin{center}
\setlength{\fboxrule}{0.8pt} 
\setlength{\fboxsep}{4pt}    

\fcolorbox{black}{gray!10}{%
  \parbox{\dimexpr\linewidth-2\fboxsep-2\fboxrule\relax}{%
    \hspace*{2pt}
    \textbf{Takeaway:} Wearable haptics can be used to provide feedback on limb positioning and body orientation.%
    \hspace*{2pt}
  }%
}
\end{center}}

\paragraph{\textbf{Off-body haptics}}
G2, G3, G4, and G7 used the provided 3D models to learn \textbf{body poses}, by either having an instructor configure the model into the pose or by positioning the model into poses themselves and receiving confirmation. G3 and G7 explored having an automated 3D model that would be pre-programmed with specific poses that learners could feel. G6 and G8 described a tactile mat with markers and vibrations. Both groups incorporated a camera to feedback on the learner's motions while the vibrating mat indicates where to move. D6 used physical anchors for instruction, having participants stand with their elbow to a wall to learn arm movements and better embody the kinesthetic metaphor in Table \ref{tab:metaphors}. 

G6 and G8 discussed off-body haptic \textbf{feedback}. G6 incorporated a textured `dance pad', where teachers may tell learners to put their feet on spots with certain textures during instruction. Similarly, G8's system involved a textured mat with indicators for orientation in degrees. G8 proposed having the mat pre-programmed with cha-cha steps. Thus, as the user completes the routine, the mat is able to record the movements and provide feedback on positioning and weight distribution. 

\aptLtoX{\begin{shaded}
\noindent \textbf{Takeaway:} Off-body haptics offer external anchors that allow pose description and provide feedback on location and orientation. 
\end{shaded}}{
\begin{center}
\setlength{\fboxrule}{0.8pt} 
\setlength{\fboxsep}{4pt}    

\fcolorbox{black}{gray!10}{%
  \parbox{\dimexpr\linewidth-2\fboxsep-2\fboxrule\relax}{%
    \hspace*{2pt}
    \textbf{Takeaway:} Off-body haptics offer external anchors that allow pose description and provide feedback on location and orientation. %
    \hspace*{2pt}
  }%
}
\end{center}}


\subsection{Multimodal Systems for Dance Instruction}
The following section describes methods participants used to combine different modalities, including creating multimodal vocabularies, including different modalities with various stages of learning, and incorporating feedback. Systems outlined by each group during the designing phase of the workshop are shown in Figure \ref{fig:group1to4} and Figure \ref{fig:group5to8}.

\subsubsection{\textbf{Providing a multimodal vocabulary.}}
Participants discussed the importance of developing a multimodal vocabulary that combine \textit{verbal vocabularies}, \textit{sound-to-movement mappings}, and \textit{haptic cue definitions}.

G1 and G5 provided movement vocabularies that matched all movement descriptions to parallel verbal names, sound cues, and haptic cue patterns (Figure~\ref{fig:group1to4}, Group 1 and Figure~\ref{fig:group5to8}, Group 5). G2, G3, G4 (Figure~\ref{fig:group1to4}), G5, G7, and G8 (Figure ~\ref{fig:group5to8}) included verbal names and sound cues in their movement vocabularies. 


\aptLtoX{\begin{shaded}
\noindent \textbf{Takeaway:} Systems should incorporate multimodal definitions into movement vocabularies.
\end{shaded}}{
\begin{center}
\setlength{\fboxrule}{0.8pt} 
\setlength{\fboxsep}{4pt}    

\fcolorbox{black}{gray!10}{%
  \parbox{\dimexpr\linewidth-2\fboxsep-2\fboxrule\relax}{%
    \hspace*{2pt}
    \textbf{Takeaway:} Systems should incorporate multimodal definitions into movement vocabularies. %
    \hspace*{2pt}
  }%
}
\end{center}}

\subsubsection{\textbf{Phases of dance instruction}}

All groups identified different \textit{phases} for teaching dance, including \textit{learning phases} (focused on teaching movement vocabularies, cue mappings, and learning routines) and \textit{practice phases} (focused on repetition and building muscle memory).

\paragraph{\textbf{Learning phase.}}
All groups emphasized having detailed verbal descriptions defining movement vocabularies during early parts of learning. B5a described: 

\begin{quote}
\textit{''If you were going to put this into three different parts, one of them is going to be the physical movement and positioning of the song. The other one is going to be talking about the differences in texture as they relate to you individually. And then a tempo and music to kind of put them all together.''}
\end{quote}

G2, G3, G4, and G5 noted that it is important to have specificity on the structure of the movements when first learning and less emphasis on feelings and textures in these early stages. Later, emotions, textures, and \textit{how} to move would be introduced with non-verbal audio once dancers have practiced \textit{what} to do. As this occurs, verbal descriptions also become less detailed. For example, after B8a described Rock Steps and Triple Steps in detail, she described the routine as: \textit{``Da da, cha cha cha''} to establish the rhythm, where \textit{da} represented rocks and \textit{cha} represented parts of the Triple Step (Figure \ref{fig:group5to8}, Group 8).

G1, G2, G6, and G8 described having ``onboarding'' during the learning phase, where the multimodal movement vocabulary is introduced and users are able to get used to sound and haptic cues. B8b noted, \textit{``If [the vocabulary] is something we have practice with, then you can associate them. And I think that has to be the first part […] what do the cues mean?''}

\paragraph{\textbf{Practice phase.}}
G2, G4, G5, and G6 outlined how the dance music should be introduced last when learning a routine. Music is therefore introduced during the practicing phase. D6 explained: 

\begin{quote}
    \textit{``In any dance class, we divorce the music from the steps and we add in the music later. First, we teach people how to do the motions and what their body positions are. Then, we see how it interacts with music.''}
\end{quote}

G1, G2, G4 (Figure~\ref{fig:group1to4}), and G7 (Figure~\ref{fig:group5to8}, Group 7) had sound and/or haptic cues transposed over the dance music during practice stages. By this point, cues would have become more 'automatic', and dancers would have internalized that certain sound cues or vibration patterns correspond to specific movement patterns. D4b described that this would help learners \textit{``understand where the motions are in the song''}. 

\paragraph{\textbf{Customized phases.}}
 G5 discussed how these phases can differ based on prior knowledge (Figure \ref{fig:group5to8}, Group 5). B5b explained, \textit{``I think it depends on your experience, because I'm coming from no experience. So the importance to me is to let me know how to move my body.''} However, more experienced dancers may prioritize expressiveness as they are already familiar with movement structure. 

\aptLtoX{\begin{shaded}
\noindent \textbf{Takeaway:} Systems should present modalities at different phases of the learning process, with verbal descriptions prioritized in early phases and non-verbal cues and music important in later, practice-heavy phases. 
\end{shaded}}{
\begin{center}
\setlength{\fboxrule}{0.8pt} 
\setlength{\fboxsep}{4pt}    

\fcolorbox{black}{gray!10}{%
  \parbox{\dimexpr\linewidth-2\fboxsep-2\fboxrule\relax}{%
    \hspace*{2pt}
    \textbf{Takeaway:} Systems should present modalities at different phases of the learning process, with verbal descriptions prioritized in early phases and non-verbal cues and music important in later, practice-heavy phases. %
    \hspace*{2pt}
  }%
}
\end{center}}

\subsubsection{\textbf{Tradeoffs Between Modalities}}

Verbal descriptions were noted by G2, G3, G4, G5, and G7 as versatile for communicating details of body mechanics and positioning, but several groups also highlighted drawbacks. G2, G3, G6, and G8 described how short instructions lacked enough information to reconstruct a movement, while lengthy descriptions were inefficient compared to other modalities. Groups emphasized the need for consistent vocabularies, since without defined terms, verbal language risked multiple interpretations. D6 pointed out that even detailed verbal instructions could still result in divergent movements, emphasizing the importance of consistency and feedback to avoid ambiguity. 

Sound cues were viewed as faster and more efficient than verbal descriptions, especially when overlaid with dance music. G1, G2, G4, and G8 described how learners could \textit{``tune out words''} (B1) but process sound cues automatically once mappings were learned. Participants expressed that sound made the experience feel more natural and engaging, with B4 explaining, \textit{``It’s nicer to dance to music.''} However, G3, G6, and G7 cautioned that mapping too many motion parameters to sound increased cognitive load. G6 concluded that simple mappings, such as pitch to height, were preferable in instructional contexts, even if they sacrificed representational accuracy.  

Haptics were described by G1, G3, G7, and G8 as efficient and immediate, with learners able to process cues such as left--right vibrations more quickly than directional words. Participants emphasized that haptic cues also build muscle memory, allowing movements to become more internalized over time (B8a). In partner dance contexts, D8b noted that tactile input is essential for coordination and connection, which G8 sought to replicate through haptic design (Figure \ref{fig:group5to8}, Group 8). Still, groups identified limits: G2 considered haptics overwhelming for contemporary dance with too many degrees of freedom, while G4 described haptic guidance as trial-and-error for complex motions. G8 also stressed cognitive limits, noting that learners could process only a few simultaneous taps before input became confusing, motivating them to restrict the number of haptic devices in their design.  

\aptLtoX{\begin{shaded}
\noindent \textbf{Takeaway:} \revised{While verbal descriptions are versatile in conveying body mechanics, they are inefficient compared to non-verbal modalities and leave room for ambiguity. However, sound and haptic cues should be simplified and provided in moderation to reduce cognitive load.}
\end{shaded}}{
\begin{center}
\setlength{\fboxrule}{0.8pt} 
\setlength{\fboxsep}{4pt}    

\fcolorbox{black}{gray!10}{%
  \parbox{\dimexpr\linewidth-2\fboxsep-2\fboxrule\relax}{%
    \hspace*{2pt}
    \textbf{Takeaway:} \revised{While verbal descriptions are versatile in conveying body mechanics, they are inefficient compared to non-verbal modalities and leave room for ambiguity. However, sound and haptic cues should be simplified and provided in moderation to reduce cognitive load.} %
    \hspace*{2pt}
  }%
}
\end{center}}

\subsubsection{\textbf{Repetition and Practice}}

All groups emphasized that repetition is important in dance education. G4's workflow (Figure~\ref{fig:group1to4}, Group 4) had dancers start from the beginning of the sequence after adding one new motion each time. B4 noted, \textit{``It helps build muscle memory to keep doing the same movements.''} Their system introduced verbal descriptions and sound effects intermittently. Dancers would learn 1-2 motions with verbal descriptions and sound cues. After new motions are added, the choreography is repeated from the beginning for reinforcement. 

\aptLtoX{\begin{shaded}
\noindent \textbf{Takeaway:} Repetition of steps is important for dance instruction systems. 
\end{shaded}}{
\begin{center}
\setlength{\fboxrule}{0.8pt} 
\setlength{\fboxsep}{4pt}    

\fcolorbox{black}{gray!10}{%
  \parbox{\dimexpr\linewidth-2\fboxsep-2\fboxrule\relax}{%
    \hspace*{2pt}
    \textbf{Takeaway:} Repetition of steps is important for dance instruction systems. %
    \hspace*{2pt}
  }%
}
\end{center}}

\subsubsection{\textbf{Customization}}

\revised{Systems offered areas of customization based on individual needs and dance styles.} G4 noted that differences in skill levels of learners may lead to differences in `correctness' criteria for feedback systems. For instance, since different users may have differences in flexibility, they determined that the system may or may not check if the dancer's hands touched the ground during the lunge step when providing feedback (Figure \ref{fig:group1to4}, Group 4). 

G5's system (Figure~\ref{fig:group5to8}, Group 5) was described as a `Choose Your Own Adventure' game where users had different settings options, including level of detail in verbal descriptions, speed of the dance, adding a metronome beat, when to include sound cues, and when to include music. For beginners, G5 concluded that it is better to have verbal descriptions only, and gradually add non-verbal elements as the user becomes more proficient. However, for advanced learners, they preferred to have non-verbal elements incorporated at the beginning to establish fluidity, expressiveness, and tempo. G5 additionally noted that people have different preferences between auditory and tactile input. \revised{Similarly, G8 concluded that less descriptiveness was required for people with more dance experience. Their system design (Figure \ref{fig:group5to8}, Group 8) involved a customization step as well, where users can customize level of detail and specific haptic cue patterns. } 

\revised{Teaching techniques and correctness criteria additionally required customization according to dance style. Social dance groups, such as G6 and G8 (Figure \ref{fig:group5to8}), emphasized correctness in timing and direction over specific movement mechanics. On the other hand, D6 explained that certain competitive styles emphasized exact movement structure. This led to differences in required cues from the system, such as sound cues conveying exact limb positions: }
\begin{quote}
\revised{\textit{``In competition ballroom dance, the quality and exactly how you do something matters. In social dance, it's as long as you and your partner are having fun and you're getting the steps in time [...] So we can have the pitch [of the sound cue] indicate the height [of the hand], but does it matter for this dance?''}}
\end{quote}

\revised{D5 additionally explained that for Contemporary dance, applying personalized styles and textures is important. Therefore, it was critical for their system to empower BLV learners to express themselves by conveying more interpretive elements, such as movement fluidity and emotions (Figure \ref{fig:group5to8}, Group 5). }

\aptLtoX{\begin{shaded}
\noindent \textbf{Takeaway:} Dance instruction systems should be customizable to specific users based on correctness criteria, level of detail, level of experience, and modality preferences.
\end{shaded}}{
\begin{center}
\setlength{\fboxrule}{0.8pt} 
\setlength{\fboxsep}{4pt}    

\fcolorbox{black}{gray!10}{%
  \parbox{\dimexpr\linewidth-2\fboxsep-2\fboxrule\relax}{%
    \hspace*{2pt}
    \textbf{Takeaway:} Dance instruction systems should be customizable to specific users based on correctness criteria, level of detail, level of experience, and modality preferences. %
    \hspace*{2pt}
  }%
}
\end{center}}

\aptLtoX{\begin{figure}
    \centering
        \includegraphics[width=\linewidth]{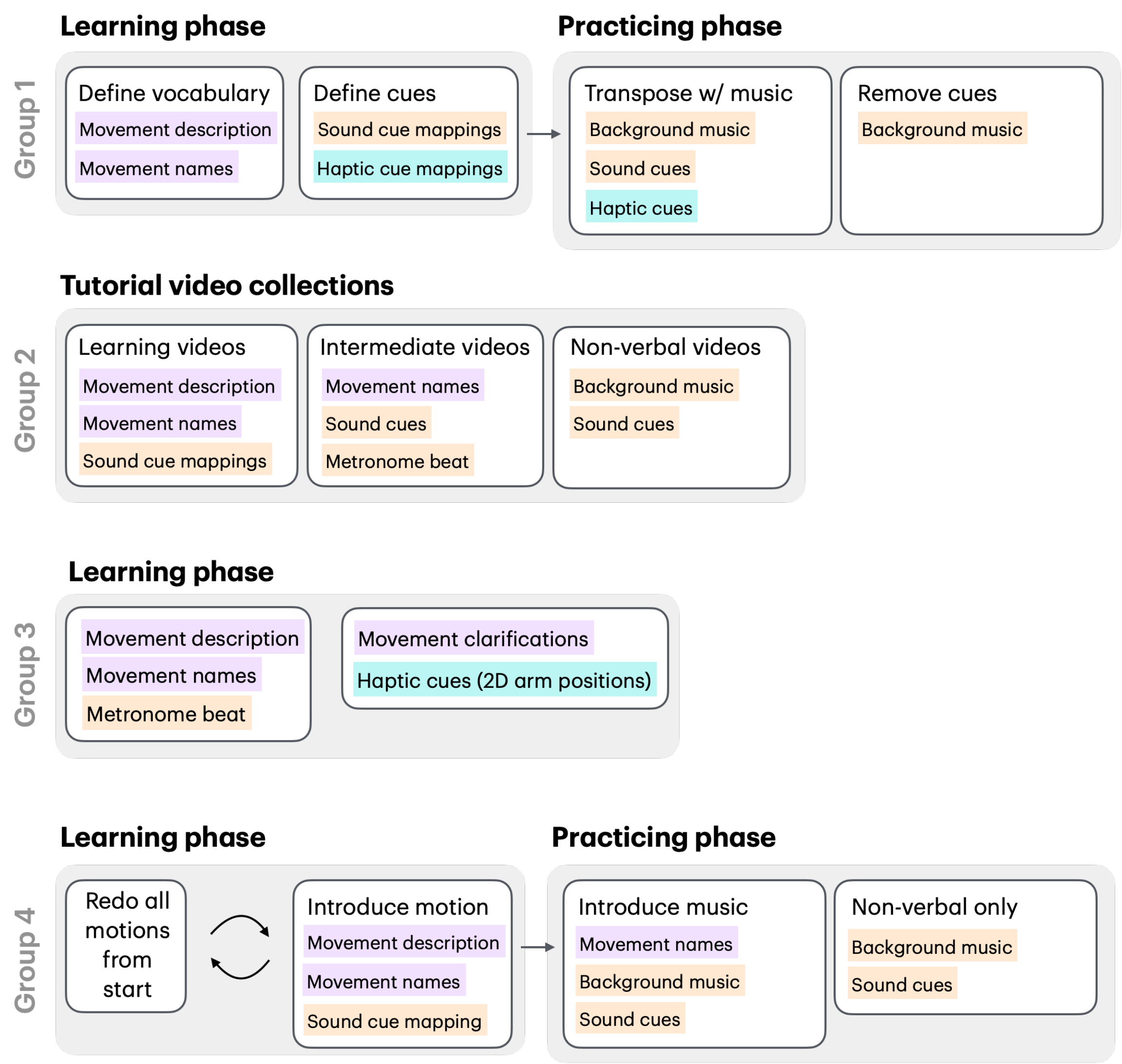}
        \caption{System design workflows for groups one to four. Groups used multiple modalities at different instructional stages, including verbal descriptions (purple, or ``movement''), non-verbal sound (orange, or ``sound'', ``music'', ``metronome''), and haptic cues (blue, or ``haptic'').}
        \label{fig:group1to4}
    \end{figure}
    \begin{figure}
        \centering
        \includegraphics[width=\linewidth]{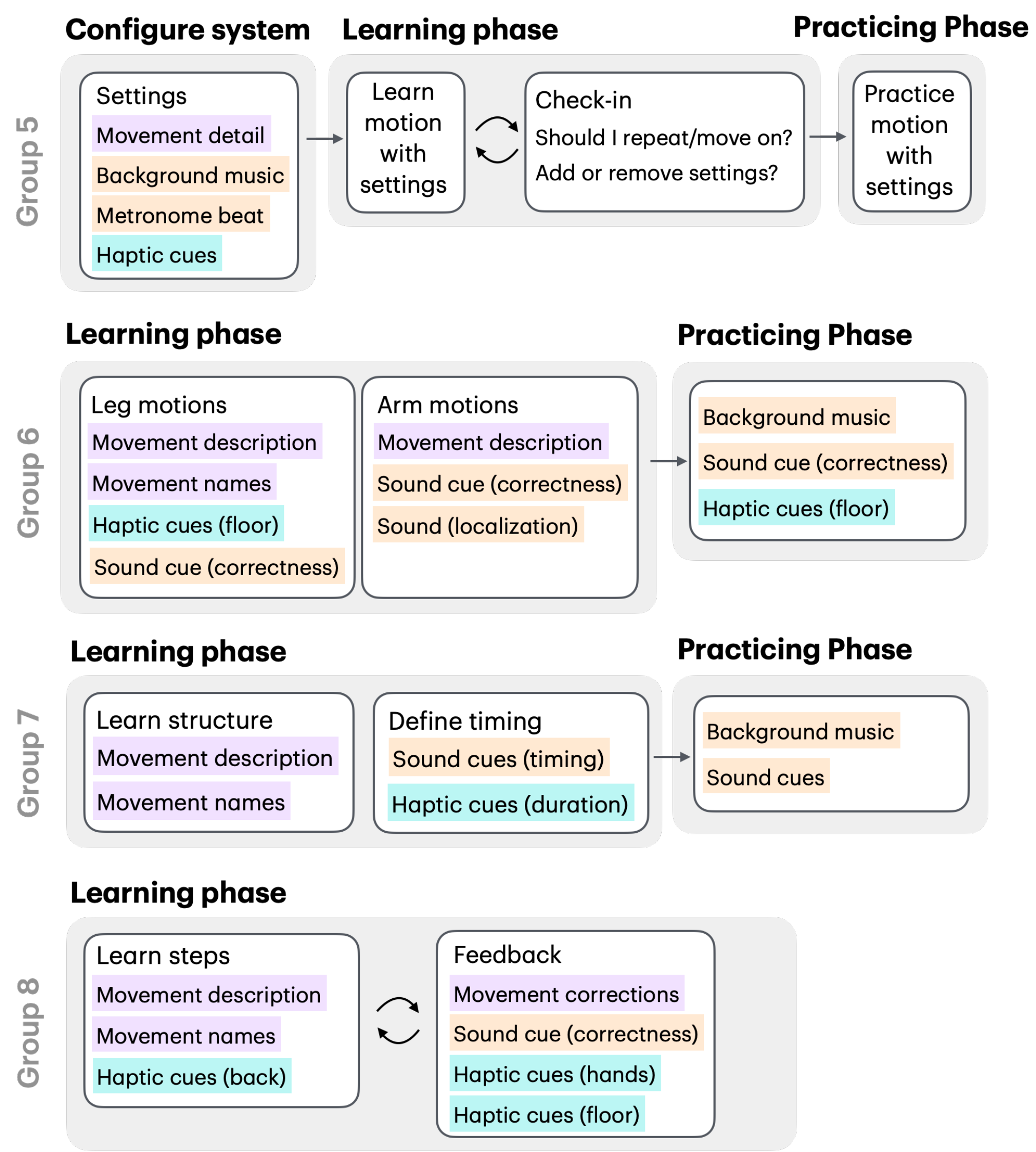}
        \caption{System design workflows for groups five through eight. Style follows Figure~\ref{fig:group1to4}.}
        \label{fig:group5to8}
\end{figure}}{
\begin{figure*}
    \centering
    \begin{minipage}{0.49\linewidth}
        \centering
        \includegraphics[width=\linewidth]{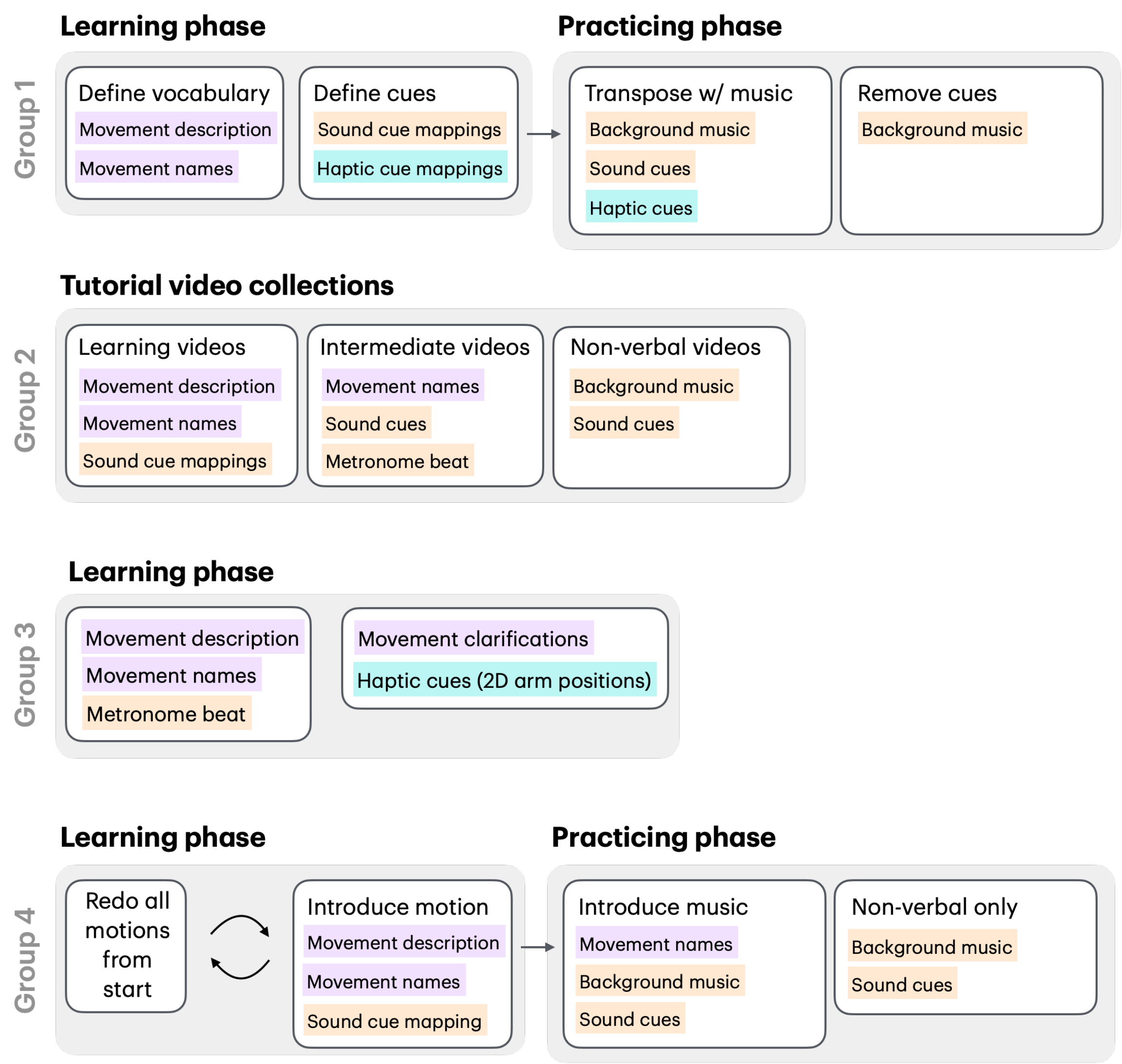}
        \caption{System design workflows for groups one to four. Groups used multiple modalities at different instructional stages, including verbal descriptions (purple, or ``movement''), non-verbal sound (orange, or ``sound'', ``music'', ``metronome''), and haptic cues (blue, or ``haptic'').}
        \label{fig:group1to4}
    \end{minipage}
    \hfill 
    \begin{minipage}{0.49\linewidth}
        \centering
        \includegraphics[width=\linewidth]{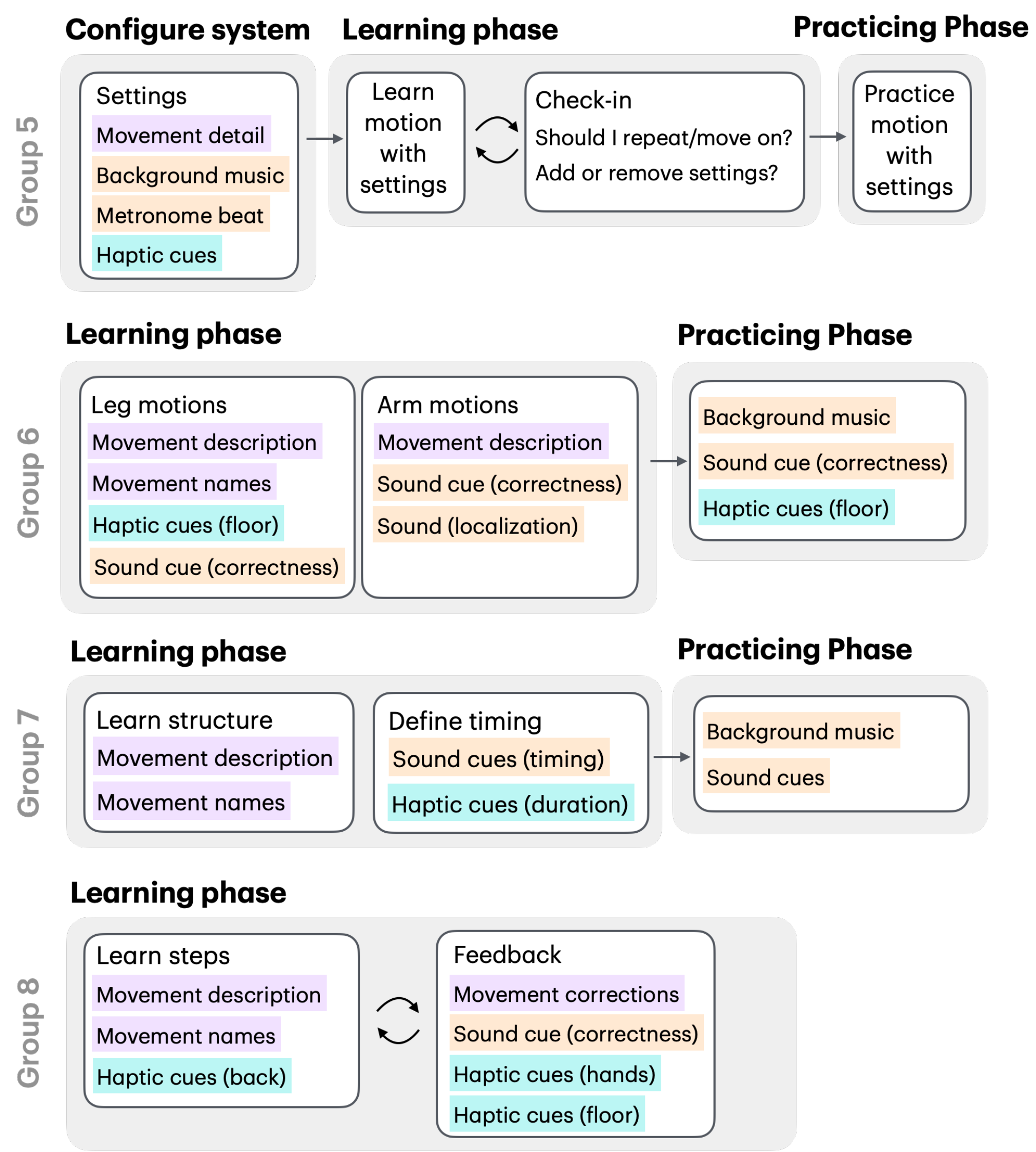}
        \caption{System design workflows for groups five through eight. Style follows Figure~\ref{fig:group1to4}.}
        \label{fig:group5to8}
    \end{minipage}
\end{figure*}}

\section{Discussion}
We synthesize design implications for potential future systems, reflect on limitations of our workshops, and propose directions for future work. 

\subsection{Design Implications}
\revised{From participants' system designs, we created 14 design implications (DI1-DI14) for future asynchronous multimodal dance instruction systems (Table~\ref{tab:design_guidelines}). These implications address five system goals: convey movement structure, convey timing and expressiveness, instruct in phases, provide feedback, and provide customization.}

\begin{table*}[h!]
\centering
\renewcommand{\arraystretch}{1.4} %
\resizebox{1.0\textwidth}{!}{%
\begin{tabular}{@{}p{4.5cm}p{1.5cm}p{1.5cm}p{7.5cm}p{2.5cm}p{2cm}@{}}
\toprule
\revised{\textbf{System Goal}} & \textbf{Implication} & \revised{\textbf{Modalities}} & \textbf{Description} & \revised{\textbf{Groups}} & \revised{\textbf{Related Work}} \\
\midrule
\revised{\textbf{Convey movement structure}} & \textbf{DI1} & \revised{V, S, H} & Create a \textit{multimodal movement vocabulary} to break down dance routines into individual, named movements. & \revised{1, 2, 3, 4, 5, 6, 7, 8} 
& \revised{\cite{Endo2024dancevideo}$^{\dagger}$,\cite{bennett2008balletlanguage}$^{\dagger}$,\cite{hassan2024dancevid}$^{\dagger}$, \cite{Blasing2015Segmentation}$^{\dagger}$,\cite{ata2019emergent}$^{\dagger}$,\cite{puri2004bharatanatyam}$^{\dagger}$} 
\\
& \textbf{DI2} & \revised{V} & \revised{Break individual movements down into verbal descriptions of specific body part motions, body parts in relation to one another, and high-level body positioning.} & \revised{2, 3, 4, 5, 8} 
& \revised{\cite{rector2013yoga}$^{*}$,\cite{desilva2025sensing}$^{*}$$^{\dagger}$,\cite{Raheb2018ACF}$^{\ddagger}$, \cite{Groff1995LabanMovement}$^{\dagger}$} \\
& \textbf{DI3} & \revised{V} & Verbally provide high-level context on dance patterns. & \revised{6, 7, 8} 
&
\revised{\cite{van2024making}$^{*}$} \\ 
& \textbf{DI4} & \revised{H} & \revised{Use haptics to indicate movement direction} & \revised{1, 6, 8} 
&
\revised{\cite{aggravi2016hapticskiing}$^{*}$,\cite{hong2017wristhaptics}$^{*}$,\cite{teng2025seeing}$^{*}$, \cite{popp2023acrosuit}$^{\ddagger}$}\\ 
\midrule
\revised{\textbf{Convey timing \& expressiveness}} & \textbf{DI5} & \revised{V, S} & \revised{Use sound-to-movement mappings and tactile metaphors to convey temporal and expressive qualities.} & \revised{1, 2, 3, 4, 5, 7} & \revised{\cite{desilva2025sensing}$^{*}$$^{\dagger}$, \cite{oppici2020sonification}$^{\ddagger}$} \\
& \textbf{DI6} & \revised{V, S} & Use non-verbal beats and verbal counts to establish rhythm. & \revised{2, 5, 8} 
&
\revised{\cite{keevallik2024repetition}$^{\dagger}$,\cite{lee2024expanding}$^{\dagger}$} \\ 
& \textbf{DI7} & \revised{H} & \revised{Use haptics to indicate movement timing.} & \revised{1, 6, 7, 8} 
&
\revised{\cite{aggravi2016hapticskiing}$^{*}$,\cite{morelli2010vitennis}$^{*}$,\cite{inbook}$^{\dagger}$, \cite{popp2023acrosuit}$^{\ddagger}$} \\ 
\midrule
\revised{\textbf{Instruct in phases}} & \textbf{DI8} & \revised{V, S, H} & Use \textit{learning phases} to let users learn movement structure and multimodal cue definitions. & \revised{1, 2, 4, 5, 6, 7, 8} 
& \revised{\cite{Ulfa2020CreativeDF}$^{*}$$^{\dagger}$,\cite{hassan2024dancevid}$^{\dagger}$}\\
& \textbf{DI9} & \revised{S, H} & Use \textit{practicing phases} once users have learned movements to allow repetition, practice timing, and incorporate dance music. & \revised{1, 2, 4, 5, 6, 7, 8} 
&
\revised{\cite{zhou2021syncup}$^{\dagger}$,\cite{Fitts1967HumanPerformance}$^{\dagger}$}\\
\midrule
\revised{\textbf{Provide feedback}} & \textbf{DI10} & \revised{H} & Use haptic and tactile elements to provide spatial and orientation feedback. & \revised{1, 3, 4, 6, 7, 8} & \revised{\cite{morelli2010vibowling}$^{*}$,\cite{seham2015extending}$^{\dagger}$} 
\\
& \textbf{DI11} & \revised{S} & Use sound cues to provide correctness feedback. & \revised{6, 8} 
& \revised{\cite{oh2013follow}$^{*}$,\cite{franccoise2016soundguides}$^{*}$}
\\
& \textbf{DI12} & \revised{V} & Include verbal check-ins and incremental corrections. & \revised{2, 4, 5, 6, 8}
& \revised{\cite{huh2025vid2coach}$^{*}$,\cite{ning2025aroma}$^{*}$}\\
\midrule
 \revised{\textbf{Provide customization}} & \textbf{DI13} & \revised{V, S, H} & Allow customization based on prior knowledge of dance and skill level. & \revised{4, 5, 6, 8} 
 & \revised{\cite{blanchet2025tiktokdancevids}$^{*}$}\\
& \textbf{DI14} & \revised{V, S, H} & \revised{Customize system based on dance style being taught.} & \revised{5, 6, 7} 
& \revised{\cite{lee2024expanding}$^{\dagger}$,\cite{conroy2025stanfordphilosophyofdance}$^{\dagger}$}
\\
\bottomrule
\end{tabular}
}
\caption{\revised{Design implications for accessible asynchronous dance instruction systems categorized into core system goals, including modalities addressed, participant groups that incorporated each implication, and related work supporting each implication. V = verbal, S = sound (non-verbal), H = haptic/tactile. Citations marked with $^{*}$ are accessibility-related, $^{\dagger}$ are dance-related, and $^{\ddagger}$ are non-accessibility work using the technique in other tasks.}}
\label{tab:design_guidelines}
\end{table*}

\subsubsection{\revised{\textbf{Convey movement structure.}}} 
An existing dance instruction approach is breaking down routines into manageable pieces, such as eight-counts~\cite{Blasing2015Segmentation}. \revised{Without access to visual demonstrations of these pieces, BLV learners relied on a \textbf{multimodal movement vocabulary} of named movements \textbf{(DI1)}. 
Future systems may draw from past work using Temporal Convolutional Networks (TCNs)~\cite{Endo2024dancevideo} and vision-language models (VLMs)~\cite{huh2025vid2coach} to segment videos into discrete steps. }

\revised{While sighted learners might rely on visual demonstrations to learn each segmented movement~\cite{blasing2018watchinglistening, Blasing2015Segmentation}, BLV participants incorporated a second level of segmentation with granular verbal descriptions of individual body parts \textbf{(DI2)} and haptic cues to indicate movement direction \textbf{(DI4)}. Future systems may explore generating detailed descriptions through established VLM-based movement understanding techniques~\cite{chen2024motionllm} and motion-to-text models~\cite{jiang2024motiongpt}. 
Such systems may draw from frameworks such as Laban Movement Analysis ~\cite{Groff1995LabanMovement}, which categorizes movements based on moving body parts, orientation/direction of these parts, and overall body shape. 
Dance styles such as Ballet, Cha-cha, and Hip-Hop contain standardized, named movements~\cite{ata2019emergent, bennett2008balletlanguage} which can be incorporated into systems for greater consistency.
Wearable haptic bracelets can additionally indicate movement directions~\cite{hong2017wristhaptics} (\textit{e.g.,} the front of your ankle vibrating meaning to move your foot forward). }

\revised{As BLV learners cannot visually preview routines, providing \textit{high-level context} (\textit{e.g., ``this dance works in opposites; every time you step forward, step backwards''}) is valuable \textbf{(DI3)}. Existing tools for accessible video descriptions~\cite{van2024making} provide high-level summaries of video content, which may be extended to dance routines. }

\subsubsection{\revised{\textbf{Convey Timing and Expresssiveness}}}
Many dances contain \textit{temporal} (time-based) and \textit{expressive} (feeling-based) qualities~\cite{smith2014dance}. 
\revised{Participants used non-verbal sound cues and metaphors to convey such aspects that are difficult to describe concisely \textbf{(DI5)}---\textit{e.g.,} the duration of the sound matching movement duration. }
\revised{Future systems may draw from existing work using sound to convey movement emotion and fluidity~\cite{frid2018motionfluidity, landry2020interactive}. Prior work on accessible dance~\cite{desilva2023understanding, blasing2021dance} additionally highlights using metaphors to convey movement qualities. Our work preliminarily found that \textit{tactile metaphors} were most commonly used for accessible descriptions. Future work may involve more formal evaluations of effective metaphor types for describing complex movement.} 

\revised{Sound beats and verbal counts to indicate rhythm \textbf{(DI6)} is common in existing instruction~\cite{keevallik2024repetition, lee2024expanding}. Our workshops showed that these techniques provide crucial grounding for BLV participants who rely on auditory information in place of visual demonstrations. Future dance instruction systems may therefore include a metronome beat and/or verbal counts as part of their movement descriptions. Participants also found haptics useful for cueing when to move \textbf{(DI7)}. Future systems can incorporate wearable haptics with vibrations timed with verbal instructions. However, such systems should consider limiting the number of haptic devices to reduce cognitive load. }

\subsubsection{\revised{\textbf{Instruct in Phases.}}}
\revised{Existing methods for motor skill learning involve gradually increasing movement autonomy with repetition and practice~\cite{hassan2024dancevid, Fitts1967HumanPerformance, Ulfa2020CreativeDF}. Our work shows that such structured learning, repetition, and practice is particularly important for allowing BLV learners to internalize spatial information that they cannot visually perceive.
Accessible asynchronous dance instruction systems may include early instructional phases that prioritize detail in movement definitions, verbal descriptions, and specific feedback as users internalize movement structure \textbf{(DI8)}. Learners can then practice with non-verbal cues to build muscle memory and establish timing and expressive qualities \textbf{(DI9)}. }
\revised{BLV participants emphasized repetition as important for developing body awareness and internalizing choreography. Future systems should thus allow for repeatable segments that users can practice \textbf{(DI9)}. }


\subsubsection{\revised{\textbf{Provide Feedback}}}
\revised{Feedback is widely used in dance instruction. However, the ambiguity of verbal descriptions is higher for BLV learners~\cite{desilva2025sensing, blasing2021dance}, while sighted learners can implicitly receive feedback through visual comparison~\cite{blasing2018watchinglistening}. BLV participants thus emphasized frequent, multimodal feedback to confirm orientation and structure and increase confidence in their movements. }

\revised{Participants found physical elements helpful for providing spatial/orientation feedback \textbf{(DI10)}. Systems may incorporate textured or haptic mats with markings to indicate orientation. Participants also found sound cues \textbf{(DI11)}, such as a bell sound when a movement is done correctly, useful for correctness feedback and encouragement. For complex movements involving multiple body parts, participants emphasized incremental verbal feedback \textbf{(DI12)}. Future systems can use pose models~\cite{fang2022alphapose} to compare learner attempts and reference videos and suggest corrections~\cite{rector2013yoga} or provide sound feedback. }



\subsubsection{\revised{\textbf{Provide Customization}}}
As outlined by participants and prior work~\cite{aflatoony2020ATmakers, jones2024customization, kane2014collaboratively}, customizable technologies improve the experience of people with disabilities. Our results showed clear differences across learners: some valued detailed auditory explanations, while others preferred tactile or haptic cues that allowed them to physically establish orientation and weight. Users additionally required varied levels of description detail. \revised{For example, participants B5a, B6, B7b and B8 had prior dancing experience, allowing them to better follow the dance to the point where they could teach other participants. Comparatively, BLV participants with no experience required more detailed instructions to learn the same dances. Asynchronous dance instruction systems should thus allow for customizable experiences that acccount for prior knowledge of dance style and skill level \textbf{(DI13)}. Our workshops additionally revealed system differences across dance styles \textbf{(DI14)}, discussed in section \ref{applyingDIs}.}


\subsection{\revised{Applying Design Implications to Asynchronous Dance Instruction Systems for BLV Learners}} \label{applyingDIs}

\revised{To support researchers and practitioners applying the co-designed system architectures and design insights in their work, we share example dance instruction systems, and describe how design implications may vary across dance styles.}

\subsubsection{Example Accessible Asynchronous Dance Instruction Systems}
\revised{Our two hypothetical examples of accessible systems each incorporate a subset of our design implications to guide future systems, derived from participant co-designed systems. One example is for a system addressing \textit{expressive dance routines}, such as with Ballet, Hip-hop, and Contemporary dance. The second example is a \textit{virtual dance partner} to teach BLV users social dance styles.}

\revised{\textbf{Expressive Dance Routine Instruction System:} Informed by G4 and G5 designing systems for Contemporary dance routines, this system example teaches a \textit{choreographed routine} (a set of predefined movements) that incorporates stylistic elements such as weight, fluidity, and expressiveness. }

\revised{In a preprocessing phase, the system takes a video of a choreographed routine and segments it into a set of individual movements \textbf{(DI1)} (\textit{e.g.}, using a Temporal Convolutional Network to segment distinct dance movements as in prior work~\cite{Endo2024dancevideo}). Next, the system provides detailed text descriptions of each movement \textbf{(DI2)} (\textit{e.g.}, with a motion-specific vision language model~\cite{chen2024motionllm}). For each segmented movement, the system also creates sound-to-movement mappings \textbf{(DI5)}. Direct mappings are created by using pose detection models such as MediaPipe~\cite{sengar2024mediapipe, lugaresi2019mediapipe} and AlphaPose~\cite{fang2022alphapose} to extract joint data and sonify them (\textit{e.g.,} mapping pitch to height or volume to speed). To create semantic mappings, the system may first generate expressive descriptions of each movement~\cite{yazdian2025motionscript} and then create expressive sound cues from these descriptions with a text-to-audio model~\cite{Huang2023texttoaudio}. During system setup, users may include information on prior dance experience to provide a customized learning experience \textbf{(DI13)} --- \textit{e.g.}, the system may substitute detailed low-level descriptions for familiar dance terms for if the participant is familiar with the dance style.}

\revised{Participants can then use the system by freely navigating via screen reader or voice command between different movements to learn each motion by listening to the verbal and sound cues \textbf{(DI8)}. Once a user is comfortable with steps, they may practice them at full speed with non-verbal cues and existing dance music during practice phases \textbf{(DI9)}. Throughout these phases, the system will further collect a video stream of the user's attempts for each step and provide verbal corrections \textbf{(DI12)} and positive sound cues once a user performs a step correctly \textbf{(DI11)}. }



\textbf{\revised{Virtual Dance Partner System:}}
\revised{Informed by G1, G6 and G8 designing systems for Salsa and Cha-cha, we envision a virtual dance partner for teaching steps from social dance videos that often demonstrate repeating standardized steps. 
During learning phases \textbf{(DI8)}, this system will cover a set of pre-programmed, standardized steps (\textit{e.g.}, rock-step, triple-step) \textbf{(DI1)} using verbal descriptions \textbf{(D12)} combined with a set of back-worn haptic devices \textbf{(DI4)}. Each step pattern will be individually defined with verbal descriptions of feet placement and pre-programmed haptic demonstrations of the step. We envision a total of four vibrotactile motors positioned horizontally across the learner's back, where vibrations indicate foot contact with the ground. Having four motors will allow the system to demonstrate steps where feet cross over one another, such as in the \textit{triple-step}. Haptic vibration intensity will correspond to weight distribution of the steps (\textit{e.g.}, a light vibration for a toe tap, a heavier vibration for weight shifts).}

\revised{The system will include wrist-worn haptics~\cite{hong2017wristhaptics} that vibrate to cue movements and indicate direction \textbf{(DI4, DI7)} (\textit{e.g.,} left wrist vibrating means to step left, a double vibration means to turn). Once learning the named movements, users are able to practice these steps with the wrist-based haptic cues, before eventually removing all cues \textbf{(DI9)}. Users are able to adjust vibration intensity for the haptic devices based on their preference, or decide to remove any set of haptics (\textit{e.g.,} no back-worn haptics during learning) \textbf{(DI13)}.}

\subsubsection{\textbf{Application Across Dance Styles}}
\revised{Dance encompasses a wide range of styles differing in movement standardization and learning techniques~\cite{conroy2025stanfordphilosophyofdance, Sparshott1988readingdancing, nicholas2004rethinking}. Certain styles emphasize standard positions and sequences (\textit{e.g.}, ballet~\cite{bennett2008balletlanguage}, Indian classical dance~\cite{puri2004bharatanatyam}). Other styles like Butoh~\cite{kasai1999butoh}, Gaga~\cite{katan2016gaga}, and Contemporary Improvisation~\cite{nakano2012improv} prioritize sensation, and internal imagery over the reproduction of fixed shapes~\cite{lepecki2006exhausting}. Instruction for these dance styles rely on sensory descriptions to guide dancers~\cite{seham2015extending}, offering more inclusive learning by de-emphasizing the reliance on visual replication. In this work, we address asynchronous learning from online dance tutorial videos, which predominantly feature structured movement routines and timing. This sequence-based, visually taught format presents distinct challenges for BLV audiences. However, we acknowledge the broader landscape of more accessible, improvisational dance practices that are less represented in such online dance tutorials.} 

By examining contrasts between groups, our workshops revealed differences in modality preferences and system designs across different dance styles \revised{\textbf{(DI14)}}. Tactile and non-verbal communication are critical aspects of partner dancing~\cite{kinoe2022analysis}, aligning with our findings where haptics was favored by all groups addressing partner dance clips \revised{\textbf{(DI4, DI7, DI10)}}. Haptic feedback is less descriptive than sounds and narration, and was therefore less suited for conveying complex movements in expressive dance instruction systems (e.g., for Contemporary dance). In such cases, sound cues and tactile metaphors \revised{\textbf{(DI5)}} were more beneficial. \revised{For dance styles with exact, pre-defined movement where precision is important (\textit{e.g.}, for Ballet), detailed verbal descriptions \textbf{(DI2-DI3)} were more critical. }

\subsection{Limitations}

Our workshop focused on multimodal technologies for asynchronous dance instruction, and the specific dance styles considered were dependent on the background of the dance teachers within each group. Although our study found significant differences in system preferences across different dance styles, we did not consider all dance styles and future work with a narrowed focus on a specific style would provide more narrowed considerations. 
\revised{Further, our insights may fail to support asynchronous learning of dance practices that are highly improvisational in which learning movements or sequences of movements from others may not be prioritized.}
Additionally, our workshops are exploratory and thus our design implications, system designs, and modality considerations are preliminary. Future work may explore empirical validation of design implications and \revised{initial low-fidelity systems through high-fidelity prototype development, testing, and refinement with co-design.} 
During our workshops, we provided participants with low-fidelity prototyping materials (\textit{e.g.}, a simple soundboard, tactile stickers to represent haptics), but more advanced haptic devices, spatial audio systems, and automated movement analysis may expand the scope of design possibilities and limitations considered by the participants. 
\revised{Furthermore, the in-person setting does not fully reflect the experience of BLV individuals learning asynchronously. Future evaluation of the design implications and proposed system workflows in remote settings would be beneficial to validate or refine our design implications.}

\subsection{Future Work}

Our work reveals future opportunities for expanding asynchronous instruction to new types of movements and learners, involving instructors in asynchronous instruction systems, and enabling long-term adaptive asynchronous instruction systems. 

\subsubsection{\textit{\textbf{Generalizability.}}} Multimodal instruction can generalize across movement practices, abilities, and learning preferences. Although our work focuses on dance, the design implications we propose extend to other domains where movement is taught through routines composed of specific poses and transitions. For example, martial arts~\cite{martialarts, jennings2020cultivation} and aerobics~\cite{sun2021aerobics} rely on choreographed sequences that combine discrete positions with fluid movements. 
Similar to dance, these practices require learners to internalize not just static poses but timing, rhythm, and fluidity. Our DIs of multimodal vocabularies and feedback could support instruction in these contexts as well. For instance, using existing or system-specific named poses could form a multimodal vocabulary for martial arts routines, while haptic cues guide stance corrections or foot placement.

Sound mappings could convey qualities such as speed, impact, or balance, paralleling how they are used to express weight and fluidity in dance.
Extending to other populations also reveals new opportunities for future work. For DeafBlind learners, systems could limit reliance on auditory feedback and emphasize using haptics to convey rhythm and expressiveness. Furthermore, the segmentation of routines and movements into clear steps may be beneficial for neurodivergent populations~\cite{stavrakiadhd, sharer2016isolating}.

\subsubsection{\textit{\textbf{Instructor-authored systems for \revised{asynchronous} instruction.}}} While future systems may be fully automated (\textit{e.g.}, converting a dance video to an instructional system, similar to Vid2Coach for cooking videos~\cite{huh2025vid2coach}), dance instructors revealed that they may be interested in making custom material for systems (\textit{e.g.}, D6 expressed wanting to re-make dance videos with useful descriptions for BLV audiences). In the future, we may explore authoring systems for instructors seeking to create accessible asynchronous instruction materials. For example, the instructors could define their own movement descriptions, names, sounds, and haptic cues that are incrementally applied by the system as students learn. In addition, with instructor involvement, future work can explore opportunities to blend synchronous and asynchronous approaches (\textit{e.g.}, generating an instructional practice system based on in-class activities).

\subsubsection{\textit{\textbf{Customizable vs. adaptive movement learning systems. }}} While our findings clearly reveal that BLV learners want to be able to customize their system experiences, this requires BLV learners knowing what options they want to use up front. In the future, we could explore closed-loop approaches to automatically adjust the descriptiveness level, modalities, or feedback type (\textit{e.g.}, positive vs. corrective) based on learner performance. Learners may be able to adjust their customization preferences, then let the system adapt within their preferences to achieve a better outcome compared to what learner-driven customization or system-driven adaptivity could do alone. To co-design such highly technical systems, future work may explore Wizard of Oz instructional systems over a longer instruction period to surface learner preferences.

\section{Conclusion}

Accessible movement education as a whole is underexplored in HCI. We conducted three co-design workshops with BLV individuals, dance instructors, and domain-specific experts (sound, haptics, AD) to understand user needs and preferences for asynchronous dance instruction systems. We identified the challenges BLV learners face when learning dance with videos, and examined how different modalities can be applied for learning and feedback. From these findings, \revised{we surfaced co-designed system architectures and developed a set of design implications (DI1–DI14) that address instruction of movement structure, timing, and expressiveness and provide feedback, practice, and customization}. From the insights gained in this work, we can inform the design of systems addressing movement education for a diverse set of populations and activities involving movement sequences, expressiveness, and precise timing. 

\begin{acks}
We thank our participants for their contribution in our workshops.  
\end{acks}

\bibliographystyle{ACM-Reference-Format}


\end{document}